\documentclass[a4paper,11pt,headings=big,DIV=12]{scrartcl}
\pdfoutput=1
\usepackage{amsmath,amssymb,mathtools}

\usepackage[usenames, dvipsnames]{color}
\usepackage{tcolorbox} 
\definecolor{mygray}{gray}{0.2}
\addtokomafont{disposition}{\rmfamily\boldmath}
\usepackage{bbold} 
\usepackage{verbatim}

\usepackage{jheppub} 

\usepackage{bm,amssymb,slashed,graphicx,multirow,soul,mathtools,xspace,array}  
\usepackage{float}                   
\allowdisplaybreaks
\usepackage{ bbold }
\usepackage{subfigure}
\usepackage{soul}
 \usepackage{slashed}
 \usepackage{acronym}
\usepackage{overpic}

\definecolor{violet}{rgb}{0.94, 0.2, 0.8}
\definecolor{lightblue}{rgb}{0.39, 0.58, 0.93} 
\definecolor{asparagus}{rgb}{0.53, 0.66, 0.42}

\acrodef{PDG}[PDG]{Particle Data Group}
\acrodef{OPE}[OPE]{Operator Product Expansion}
\acrodef{FCNC}[FCNC]{flavour-changing neutral current}
\acrodef{RHC}[RHC]{right-handed currents}
\acrodef{SM}[SM]{Standard Model}
\acrodef{NP}[NP]{New Physics}
\acrodef{MFV}[MFV]{Minimal Flavour Violation}
\acrodef{SD}[SD]{short-distance}
\acrodef{LD}[LD]{long-distance}
\acrodef{DA}[DA]{distribution amplitude}

\newcommand{\ellint}{10e}

\newcommand{\matel}[3]{\langle #1|#2|#3\rangle}
\newcommand{\al}{\alpha}
\newcommand{\be}{\beta}
\newcommand{\ga}{\gamma}
\newcommand{\de}{\delta}
\newcommand{\la}{\lambda}

\newcommand{\GeV}{\,\mbox{GeV}}
\newcommand{\MeV}{\,\mbox{MeV}}

\newcommand{\Cdot}{\!\cdot \!}
\newcommand{\mi}{\!-\!}
\newcommand{\pl}{\!+\!}

\newcommand{\TAB}{Tab.~}
\newcommand{\FIG}{Fig.~}
\newcommand{\FIGs}{Figs.~}
\newcommand{\SEC}{Sec.~}
\newcommand{\SECs}{Secs.~}
\newcommand{\APP}{App.~}
\newcommand{\APPs}{Apps.~}
\newcommand{\EQ}{Eq.~}
\newcommand{\EQs}{Eqs.~}

\newcommand{\Ima}{\textrm{Im}}
\newcommand{\disc}{\textrm{disc}}

\newcommand{\mlone}{m_{\ell_1}}

\newcommand{\muhc}{\mu_{\text{hc}}}

\newcommand{\FRtwo}{(2)}
\newcommand{\FRthree}{(3)}

\newcommand{\pBbar}{\bar{p}_B}

\newcommand{\mga}{m_\gamma}

\newcommand{\CV}{C_V}
\newcommand{\CA}{C_A}

\newcommand{\Tg}{\theta_{\ga}}
\newcommand{\Tl}{\theta_{\ell}}

\newcommand{\qz}{q_0}
\newcommand{\qzh}{\hat{q}_0}

\newcommand{\Tlz}{\theta_0}

\newcommand{\cl}{c_{\ell}}

\newcommand{\clz}{c_{0}}

\newcommand{\Fg}{\phi_{\ga}}

\newcommand{\lonetwo}{\ell_{1,2}}
\newcommand{\ltwoone}{\ell_{2,1}}
\newcommand{\lone}{\ell_1}
\newcommand{\ltwo}{\ell_2}

\newcommand{\bltwo}{\bar{\ell}_2}
\newcommand{\pK}{p_K}

\newcommand{\modveclone}[1]{|\vec{\ell}_1^{#1}|}

\newcommand{\des}{\de_{\textrm{ex}}}

\newcommand{\TT}{\ell}

\newcommand{\Bin}{{\bar{B}}}
\newcommand{\Kout}{{\bar{K}}} 

\newcommand{\LO}{{\textrm{LO}}}

\newcommand{\ORD}{{\cal O}}

\definecolor{violet}{rgb}{0.94, 0.2, 0.8}
\definecolor{lightblue}{rgb}{0.39, 0.58, 0.93} 
\definecolor{lightgreen}{rgb}{0.1, 0.73, 0.33}

\newcommand{\PHOTOS}{\texttt{PHOTOS}\xspace}
\newcommand{\PHOTONS}{\texttt{PHOTONS++}\xspace}
\newcommand{\SHERPA}{\texttt{SHERPA}\xspace}

\newcommand{\mBrec}{m_{B}^{\textrm{rec}}}
\newcommand{\mBrecl}{m_{B}^{\textrm{rec} ,\ell}}
\newcommand{\cutter}{ \Delta \omega^2}

\newcommand{\fth}{f^{\mathrm{th}}}
\newcommand{\egammacut}[1]{E_{\gamma\, ,\mathrm{cut}}^{(#1)}}
\newcommand{\egamma}[1]{E_{\gamma}^{(#1)}}
\newcommand{\qnotsq}{\mbox{\ensuremath{q_{0}^{2}}}\xspace}
\newcommand{\qsq}{\mbox{\ensuremath{q^{2}}}\xspace}

\newcommand{\ctl}{\mbox{\ensuremath{c_{\ell}}\xspace}}
\newcommand{\ctg}{\mbox{\ensuremath{c_{\gamma}}\xspace}}

\newcommand{\BKll}{\mbox{\ensuremath{\bar B^0 \to \bar K^0   \ell^+\ell^-}}\xspace}

\newcommand{\JPsi}{\mbox{\ensuremath{J/\Psi}}\xspace}

\DeclareOldFontCommand{\tt}{\normalfont\ttfamily}{\mathtt}

\newcommand*{\mathcolor}{}
\def\mathcolor#1#{\mathcoloraux{#1}}
\newcommand*{\mathcoloraux}[3]{%
  \protect\leavevmode
  \begingroup
    \color#1{#2}#3%
  \endgroup
}

\title{\boldmath  
QED in $\bar B \to \bar K  \ell^+\ell^-$ LFU ratios: \\ 
Theory versus Experiment, a Monte Carlo Study
}

\author[1]{Gino Isidori,}
\author[1]{Davide Lancierini,}
\author[1,2]{Saad Nabeebaccus,}
\author[3]{Roman Zwicky,}

\affiliation[1]{Department of Physics, Universit\"at Z\"urich, 
Winterthurerstr. 190,  CH-8057 Z\"urich, Switzerland}
\affiliation[2]{Universit\'e Paris-Saclay, CNRS, IJCLab, 91405 Orsay, France}
\affiliation[3]{Higgs Centre for Theoretical Physics, School of Physics and Astronomy, University of Edinburgh, 
Peter Guthrie Tait Road, King's Buildings, Edinburgh EH9 3FD, Scotland, UK}
\emailAdd{isidori@physik.uzh.ch}
\emailAdd{davide.lancierini@uzh.ch}
\emailAdd{saad.nabeebaccus@ijclab.in2p3.fr}
\emailAdd{roman.zwicky@ed.ac.uk}

\abstract{
Using analytic results obtained in a meson effective theory, that includes all infrared sensitive logs, we build a dedicated Monte Carlo framework to describe QED corrections in $\bar B \to \bar K \ell^+\ell^-$ for a generic form factor. For the neutral mode $\bar B^0 \to \bar K^0 \ell^+\ell^-$, we perform a detailed numerical comparison versus those obtained with the general-purpose photon-shower tool PHOTOS. The comparison indicates a good agreement, at the few per-mil level, when focusing on the rare mode only. In addition, our framework allows us to investigate the impact of the charmonium resonances. Interference effects, not described by PHOTOS in the experimental analysis, are found to be small in the dilepton invariant mass region $q^2 < 6 \textrm{GeV}^2$, which is used to determine $R_{K^{(*)}}$. Using a semi-analytic framework we assess the full, rare and resonant, mode. Based thereupon, we discuss strategies to check the subtraction of the resonant mode, which has a sizeable impact at $q^2 \approx 6 \textrm{GeV}^2$ in the electron mode.
}

\begin{document}


\maketitle

\flushbottom

\setcounter{tocdepth}{3}
\setcounter{page}{1}
\pagestyle{plain}

\section{Introduction}

Within the Standard Model (SM), the Yukawa coupling is the only interaction that distinguishes the different 
fermion families. In the lepton sector, all the Yukawa couplings are small compared to the SM gauge couplings, 
giving rise to an approximate accidental symmetry known as  Lepton Flavour Universality (LFU). 
This symmetry holds to a very good accuracy within the SM, especially for the two lightest families
($e$ and $\mu$), and it can be tested to high accuracy in $B$ meson decays, where the kinematic effects due to light 
lepton masses are small (see e.g. Ref.~\cite{AIS,Bifani:2018zmi} for a review).

Particularly interesting in this respect are the $\mu/e$ LFU ratios in 
flavour changing neutral currents (FCNC)  transitions~\cite{Hiller:2003js}, such as 
\begin{equation}
\label{eq:RK}
R_K|_{q_0^2 \in [q_1^2,q_2^2]{\small \GeV}^2} = 
\left. \frac{\Gamma[\Bin \to \Kout \mu^+ \mu^-]}{\Gamma[\Bin \to \Kout e^+ e^-]} \right|_{q_0^2 \in [q_1^2,q_2^2]{\small \GeV}^2}\;,
\end{equation}
where $q_0^2 \equiv (p_B - p_K)^2$. 
 In the SM, $R_K^{\textrm{SM}} \approx 1$ up to QED corrections~\cite{BIP16,Isidori:2020acz}. 
 The current experimental determination is~\cite{LHCb:2014vgu,LHCb:2019hip,LHCb:2021trn} 
\begin{equation}
\label{eq:RKex}
R_K|_{q_0^2 \in [1.1,6]{\small \GeV}^2} = 0.846^{+ 0.042 + 0.016}_{- 0.039 - 0.012}\;,
\end{equation}
and exhibits a statistically significant deviation from the theory prediction.
Similar tensions between data and SM predictions, albeit with smaller statistical significance, 
have been reported in the analogous quantities $R_{K^{*0}}$~\cite{LHCb:2017avl}, $R_{K^{*+}}$ and $R_{K_S}$~\cite{LHCb:2021lvy}.

In this paper, we assess the robustness of the theoretical determination of $R_K$ 
with respect to QED corrections, which provide the dominant source of LFU violation within the SM.
While QED corrections are tiny for fully inclusive observables (when differential in collinear-safe variables), they induce non-universal
corrections of the type $(\al/\pi) \ln (m_\ell / m_B)$ which can reach the $10\%$ level 
in the electron mode, when accompanied by tight cuts on the photon energy~\cite{BIP16,Isidori:2020acz}.
These effects are corrected for by the experimental collaborations: the value in Eq.~(\ref{eq:RKex}),
as well as the results for the LFU ratios reported in \cite{LHCb:2017avl,LHCb:2021lvy},
correspond to photon-inclusive observables  (in the collinear safe differential variable  $q_0^2$,   cf.~\SEC\ref{sec:generalities}.)
However, what is really measured are 
not photon-inclusive observables: tight cuts on reconstructed $B$ mass are employed 
to reduce, amongst the different background contributions, events originated from resonant modes,
e.g. $\Bin \to \Kout (J/\Psi\to \ell^+ \ell^-)$ that leak into the signal region.  The photon-inclusive results are obtained 
 by comparing with  appropriate Monte Carlo (MC) simulations. The purpose of this paper 
is to check this procedure using a dedicated MC-framework developed on grounds on our earlier 
work~\cite{Isidori:2020acz}.  The latter  consists  of a complete differential description of $O(\alpha)$ QED 
corrections in $\bar B \to \bar K \ell^+\ell^-(\gamma)$ based on an effective meson theory. 

In the experimental analyses, QED corrections are implemented via photon shower algorithms such as \PHOTOS  
\cite{Barberio:1990ms,Barberio:1993qi,Golonka:2005pn,PHOTOS}, or the \PHOTONS  module \cite{Schonherr:2008av} of \SHERPA \cite{Sherpa:2019gpd},
where mesons are treated as point-like particles.
In \cite{Isidori:2020acz}, using gauge invariance, it was shown that no further lepton non-universal collinear logs (i.e.~$\ln (m_\ell)$ terms) 
are generated by 
structure dependent corrections, i.e.~that the point-like approximation for the mesons is a very good approximation, especially
when considering LFU ratios. 
The photon shower algorithms used by the experiments therefore do provide a very good starting point to describe data. 
In practice,  QED corrections in $\bar B \to \bar K \ell^+\ell^-(\gamma)$ are not treated perfectly
 due to  the resonant mode being simulated separately from the rare mode, therefore neglecting the respective interference. 
The latter is a potentially dangerous effect due 
to the {\em migration} towards lower $q^2$-values of events with  on-shell charmonium resonances and 
sizeable photon-energy emission: an effect which is particularly pronounced for the electron mode~\cite{BIP16,Isidori:2020acz}. We note that in the inclusive case these effects have been 
investigated in the  factorisation approximation in \cite{Huber:2020vup}, whereas we can go beyond 
since 
the $\bar B \to \bar K \Psi$ branching fractions  are known from experiment.

This paper consists of two parts. 
Firstly, the description of our  MC-framework based on~\cite{Isidori:2020acz}, 
and its comparison with~\PHOTOS  at the fully differential level, considering the rare mode only
(i.e.~the short-distance (SD) part of the decay amplitude) is discussed.
Second, going beyond the \PHOTOS analysis, we assess the impact of the charmonium resonances (or  the long-distance (LD) contribution to the decay 
amplitude), which is particularly relevant in the electron mode~\cite{LHCb:2014vgu,LHCb:2019hip,LHCb:2021trn}.
This second part is addressed in a twofold manner: i) by means of our MC-framework, assessing the impact  of the 
 SD--LD interference effects (not included in \PHOTOS), focusing on the region $ q^2 \in [1.1,6]\GeV^2$;
 ii)  by means of a semi-analytic approach, using the splitting function, assessing the complete impact of the resonant modes beyond interference terms (and resumming the leading collinear logs).

We limit our analysis to the neutral mode  $\bar B^0 \to \bar K^0 \ell^+\ell^-$. This choice does not limit the validity of our conclusions on 
SD--LD  interference effects due to charmonium resonances,  whilst it has an important simplification for the numerical study. 
In this case, we can analyse in full generality  the impact of the  hadronic form factor, without resorting to  
a derivative expansion in the underlying meson effective theory~\cite{Isidori:2020acz}.

It is noted that a first comparison of analytical estimates of QED corrections and \PHOTOS has been presented in 
Ref.~\cite{BIP16}. The present study provides significant improvements compared to~\cite{BIP16} on various aspects: 
i)~building a dedicated MC to simulate $\bar B \to \bar K \ell^+\ell^-(\gamma)$ events, 
we are able to perform an extensive study of the tool used to interpret data at a fully differential level; 
ii)~our MC is valid for generic photon kinematics, while the analysis of~\cite{BIP16} implicitly assumed tight cuts on the photon-angle emission (cf.\,\APP A.2~\cite{Isidori:2020acz});
iii)~we perform a detailed study of the effects of the resonances, taking into account also the variation of the strong phase 
between SD and LD contributions.\footnote{~Some other studies related to Monte 
Carlo are for semileptonic modes  $B \to \pi \ell \nu$  \cite{Bernlochner:2010fc} and 
$B \to D \ell \nu$ \cite{Cali:2019nwp} and are different in that they do not contain resonances from 
the phenomenological viewpoint alone.}

The paper is organised as follows. In \SEC\ref{sec:MCframe}, we introduce the the basic kinematics of the process and  the 
principles of our MC-approach. In the following \SEC\ref{sec:comparison}, we compare kinematic distributions 
obtained with our MC-simulation with those obtained with \PHOTOS. The impact of charmonium resonances 
is discussed in \SEC\ref{sec:charm}.  Finally, in \SEC\ref{sec:conclusions}, we summarise our results and present a brief 
outlook.  Technical details are deferred to \APPs\ref{app:kin}, \ref{app:charm}  to \ref{app:fthtable} and  supplementary plots are collected  in \APP\ref{app:suppplots}.

\section{Monte Carlo Framework} 
\label{sec:MCframe}

\subsection{Generalities}
\label{sec:generalities}

The process of interest  is
\begin{equation}
\Bin(p_B) \to  \Kout(\pK)   \lone(\lone)   \bltwo(\ltwo) +  \ga(k)~.
\end{equation}
In the absence of photon emission, it is
a 3-body process, while in the presence of real photon emission, it corresponds to a 
4-body process. The latter is characterised by five independent kinematic variables, cf.~\APP\ref{app:kin}.
The two kinematic variables adopted to describe the 3-body kinematics, or even the 4-body one 
 if the photon is not detected,  are 
\begin{equation} 
\label{eq:q2}
q^2 = (\lone+\ltwo)^2 \qquad\textrm{and}\qquad  c_\ell \equiv \cos \theta_\ell =  - \left(\frac{\vec{\lone}\cdot \vec{p}_K}{ |\vec{\lone}| | \vec{p}_K| } \right)_{q\textrm{-RF}}~,
\end{equation}
where $q\textrm{-RF}$ denotes the dilepton rest frame (RF).
If the $B$ momentum is known (e.g.~at the generator level, or in a $B$-factory type experimental setup)
 the following the kinematic variables
\begin{equation} 
\label{eq:q02}
q_0^2 = (p_B-\pK)^2 \qquad\textrm{and}\qquad  \clz  \equiv \cos \Tlz   =  - \left(\frac{\vec{\lone}\cdot \vec{p}_K}{ |\vec{\lone}| | \vec{p}_K| } \right)_{q_0\textrm{-RF}}~ \;,
\end{equation}
are more useful \cite{Isidori:2020acz}.
Furthermore we define 
\begin{align}
\pBbar &\equiv p_B - k = \lone + \ltwo + \pK \;,\quad  \pBbar^2= (\mBrec)^2\;,
\end{align}
which corresponds to the reconstructed $B$-meson mass 
from its visible decay products, and the variable $\des$,
\begin{equation} 
\label{eq:des}
(\mBrec)^2 = m_B^2(1 - \des) \,,  \qquad 0 < \des < 1\,,
\end{equation} 
which provides a natural choice for the physical cut-off regulating soft divergences of real photons emission. Soft and soft-collinear logs then manifest as $\ln \des$ and $\ln \des \ln m_\ell$ terms. 
Single $\ln m_\ell$ terms are referred to (hard)-collinear logs throughout; a terminology which differs at times from the ones used in soft-collinear effective theories \cite{Becher:2014oda}.

As stated in many textbooks, a photon energy cut-off, $\mBrec$ or $\des$ in our case, is sufficient to define  
IR-safe observables (for massive charged particles).
However,  this is not the procedure applied in many of today's experiments,
especially at hadron colliders.
 In this case  the event distributions are {\em fitted}  in a given window of $\mBrec  >  \mBrecl$ 
 and $\mBrec$ becomes a key differential variable. 
 Using the simulated shape in $\mBrec$, by a MC-tool (e.g. \PHOTOS), the theoretical non-radiative rate 
 in the IR-safe  differential variable  $q_0^2$ 
 is reconstructed.\footnote{In terms of the $ \qz^2 $-variable, the single-differential non-radiative rate is equivalent to the  fully photon-inclusive  
 rate up to   $\ORD(\frac{\al}{\pi})$ corrections.}   
Checking the validity of this procedure requires the  comparison of  the MC-tool
used in the data analysis with one based on a  QED calculation defined in a full theoretical framework, 
such as the one presented in~\cite{Isidori:2020acz}. 
The validation of the procedure ensures that, within the SM, the measured $R_K$ is then 
$R_K|^{\textrm{SM}}_{\textrm{reconstr.}} = 1 + \ORD(\frac{\al}{\pi})$, where $\al = e^2/(4 \pi) \approx 1/137$ is the fine structure constant.

\subsection{Basic strategy of the Monte Carlo approach}
\label{sec:basics}

The strategy of our MC-framework is based on the following steps:
\begin{itemize}
\item  We introduce a technical cut-off $\egammacut{i}$ 
on the photon energy in a given RF (indicated by the superscript $i$).
 This cut-off is chosen well below the experimental resolution on the missing energy, 
such that events with $\egamma{i}<\egammacut{i}$ can be simulated according to 3-body kinematics, while 
events with $\egamma{i}> \egammacut{i}$ are simulated according to 4-body kinematics.
\item
The 3-body and 4-body events are simulated according to the corresponding (Born-level) distributions 
reported in Ref.~\cite{Isidori:2020acz}, which depends on the  $f_{\pm}(q^2)$ 
hadronic form factors for ${B\to K}$.
The relative normalisation between 3-body events ($N_3$) and 
4-body events ($N_4$), namely the ratio
\begin{equation}
\fth  \equiv \frac{N_3}{N_4} \equiv  \frac{\Gamma_{3}}{\Gamma_{4}} =  f(\egammacut{i})~,
\end{equation}
 is the key theory input for the numerical simulation. 
 \item
 The 3-body rate is computed at $O(\alpha)$, taking into account both virtual and real corrections.
 By construction, $\Gamma_{3} $ is free from soft divergences, but it depends logarithmically on the (artificial) 
photon energy cut-off $\egammacut{i}$. It can be decomposed as 
\begin{align}
	\Gamma_{3} =\Gamma_{\mathrm{soft-log}}\ln \egammacut{i} + \Gamma_{\mathrm{rest}}^{(i)}\;.
\end{align}
Here, $\Gamma_{\mathrm{soft-log}}$ is the well-known universal (Lorentz-invariant) coefficient of the soft singularities~\cite{Weinberg:1965nx}, while 
$\Gamma_{\mathrm{rest}}^{(i)}$ is a frame-dependent  quantity, indicated by the superscript~\cite{Isidori:2020acz}. 
 \item
 A key simplification for the determination of $\fth$  is the observation that the total rate $\Gamma_{\mathrm{tot}}$ 
 is equal to the tree-level rate, $\Gamma_{\mathrm{tree}}$, up to finite (non-log enhanced) corrections of $\ORD(\alpha)$:
\begin{equation}
\Gamma_{\mathrm{tot}}\equiv \Gamma_{3}+\Gamma_{4}=\Gamma_{\mathrm{tree}}\times \left[ 1+\ORD(\alpha) \right]\;.
\label{eq:Collsafe}
\end{equation}
Neglecting the tiny $\ORD(\alpha)$ terms, this allows us to extract $\fth$ simply by the ratio $\Gamma_{\mathrm{tree}}/\Gamma_{\mathrm{3}}$,
via the relation
\begin{align}
	\label{eq:simplefth}
	\fth=\left( {\frac{\Gamma_{\mathrm{tree}}}{\Gamma_{\mathrm{3}}}-1} \right)^{-1}\;.
\end{align}
 \item
As demonstrated in~\cite{Isidori:2020acz},  the $q_0^2$ single-differential spectrum is also free from soft and collinear divergences.
This implies that the relation (\ref{eq:Collsafe}) holds not only for the total rate, but also for the $q_0^2$ single-differential rate
(or partial rates defined on a given $q_0^2$ interval). 
Using  \EQ\eqref{eq:simplefth} simplifies the numerical analysis considerably, since $\Gamma_{\mathrm{tree}}/\Gamma_{\mathrm{3}}$ can be determined 
using only 3-body phase-space integrations. The values  of $\Gamma_{\mathrm{tree}}/\Gamma_{\mathrm{3}}$ 
computed using the analytic code from \cite{Isidori:2020acz} 
relevant to the present study are reported in \TAB\ref{tab:fth} in \APP\ref{app:fthtable}.
\end{itemize}

\subsection{Numerical procedure}
\label{sec:numprocedure}

The 3- and 4-body decay rates are implemented in a numerical framework by means of the zfit package \cite{Eschle:2019jmu}. They are interpreted as (non-negative) probability distribution functions (PDFs) with an a priori unknown normalisation. 
This allows us to generate  the MC samples by means of the hit-or-miss algorithm. 
The concrete sampling  procedure, for  both $3$- and $4$-body decays,   is outlined as follows:

\begin{enumerate}
	\item A single  point in phase space, denoted by $\vec{x}$, is uniformly sampled in the kinematically  allowed region of $(\qsq,\cl)$ or $(\qsq , \pBbar^{2}, \cl,\ctg ,\phi_\gamma)$ for simulated 3- or 4-body events, respectively.  
		
	\item Using the kinematic decomposition reported in \APP\ref{app:kin},	the sampled variables are translated to the corresponding momenta of the $B$ decay products. The scalar products that enter the decay widths, as well as the decay width $\Gamma(\vec{x})$ itself, are evaluated at the sampled point in phase space.
 
	\item  A random number $r$ is extracted uniformly in the range $r\in[0,m]$ , where $m$ is the maximum value of the decay width in the allowed kinematic range.
	If $r> \Gamma(\vec{x})$ (``miss") points 1. and 2. are repeated until for one sampled $r$, $r<\Gamma(\vec{x})$ (``hit") and the point $\vec{x}$ is kept.
\end{enumerate}	
 The standard hit-or-miss algorithm can suffer from very low sampling efficiency in the case where the decay width exhibits pronounced peaks since it only accepts a fraction of extractions equal to the ratio of the volume under $\Gamma(\vec{x})$ and the volume of the hypercube containing $\Gamma(\vec{x})$ itself. This is valid in particular for the 4-body decay width $\Gamma_{4}(\vec{x})$ which is peaked close to collinear and soft regions.  In order to increase the sampling efficiency, the ``importance sampling" technique is employed. This technique consists in dividing the support, over which $\vec{x}$ is sampled, such that in each subinterval, $\Gamma(\vec{x})$ has a smaller variation than in the overall range, hence allowing to increase the sampling efficiency by one to two orders of magnitude.

\paragraph{ }
We compare our MC-approach against the EvtGen \cite{Ryd:2005zz} 
 + \PHOTOS \cite{Barberio:1990ms,Barberio:1993qi} which are the simulation software packages used by the LHCb collaboration. 
 The former is an event generator specifically designed for $B$-physics in which the decay amplitudes (models), instead of PDFs, are used for the simulation of heavy meson decays. \PHOTOS encodes the QED radiative corrections to such decays and uses a splitting function approach iteratively. In principle, this achieves the inclusion of the leading logs and thus, remaining discrepancies can be expected to be of $\ORD(\frac{\al}{\pi})$. Since 2005, when multi-photon radiation was introduced \cite{Golonka:2005pn}, there were no further public upgrades of the 
program until 2010, when \PHOTOS was moved to a $\texttt{C++}$ environment allowing the use of event records such as HepMC \cite{Dobbs:2001ck}. 
We employ version 3.64  
which enables the use of \PHOTOS in the case where there are no parent particle(s) or incoming beams generating the decaying particle, which, paired with the EvtGen package, allows us for a direct comparison with our MC simulation. In order to match the EvtGen model with our description of the decay, $\ORD(\alpha_{s})$ two-loop virtual corrections \cite{Asatrian:2001de} 
to the decay width are switched off from the default EvtGen configuration. 

Moreover, since our MC accounts for QED corrections up to $\ORD(\al)$, for each $B$ decay in which more than a real photon is emitted, only the hardest emission is considered and all the softer emissions' momenta are summed and saved into one ``particle" for further cross-checks.

\section{Direct Comparison with \PHOTOS at the Short Distance Level }
\label{sec:comparison}

\subsection{Parameterisation of the short distance amplitude}

In this section, we compare our MC-method, as described in Sec. \ref{sec:numprocedure}, 
with the \PHOTOS framework at the level of the SD contribution.    
Since both frameworks are expected to capture the leading logs one should expect differences
 to be of order of $\ORD(\frac{\al}{\pi})$ only.  Let us first define the SD amplitude. 
Following our previous 
conventions \cite{Isidori:2020acz},  we write
\begin{equation}
{\cal A}_{\Bin \to \Kout\ell^+ \ell^-} \equiv \matel{\Kout \ell^+ \ell^-  }{ (- {\cal L}_{\textrm{int}}) }{\bar{B}} = 
\frac{G_F}{\sqrt{2}} V^*_{\text{ ts}}V_{\text{tb}} \,  L_0 \Cdot H_0 + \ORD(\al)  \;,
\end{equation}
 where $L_0$ and $H_0$ correspond to the leptonic and hadronic parts which read  
 \begin{alignat}{2}
 \label{eq:L0}
&  L_0^\mu(q^2)  &\;=\;&  \bar{u}(\ell^-)\ga^\mu( \CV + \CA \ga_5)   v(\ell^+) \;, \nonumber  \\[0.1cm]
& H_0^\mu(q^2)  &\;=\;&  f_+(q^2) (p_B\pl p_K)^\mu + f_-(q^2) (p_B\mi p_K)^\mu   \;,
 \end{alignat} 
with $C_{V(A)}  =  -\alpha  C_{9(10)}/(2\pi)$, thereby neglecting the dipole operator $O_7$ as justified for a scalar meson final state.\footnote{~In the relation of $C_{V(A)}$ 
and $C_{9(10)}$ we correct a factor minus two w.r.t. the published version of \cite{Isidori:2020acz} which, however, has no impact on the result of that paper as all results are relative.}
The SD contribution consists of the standard form factors  $f_{\pm}$ . Note that  when $f_-$ is traded for the scalar form factor   $ f_0 =  f_+ + \frac{q^2}{m_B^2-m_K^2}  f_-$ 
only  $f_+$ enters the vector part $C_V \propto C_9$.  
The specific form factors are taken from \cite{Ball:2004ye} (with set 2), which is a   light-cone sum rules computation up to NLO twist-3 and $\ORD(\al_s)$, and is also used by the LHCb collaboration.

\subsection{Comparison of our Monte Carlo with \PHOTOS}

\begin{figure}[!htbp]
   \begin{center}
      \begin{overpic}[width=0.48\linewidth]{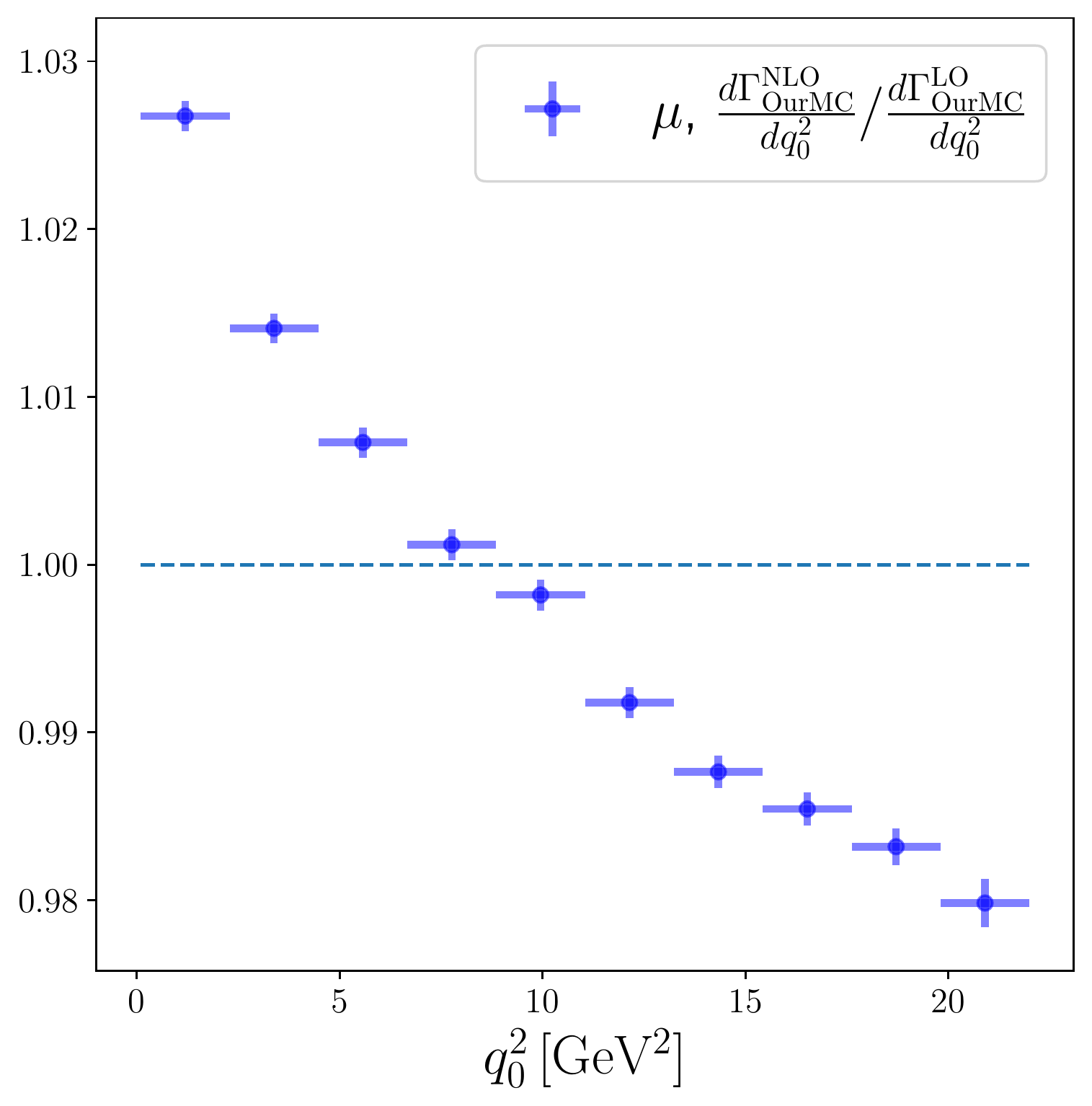}
      \put(52, 75){$\mBrec > 5.18 \GeV$}
      \end{overpic}
      \begin{overpic}[width=0.48\linewidth]{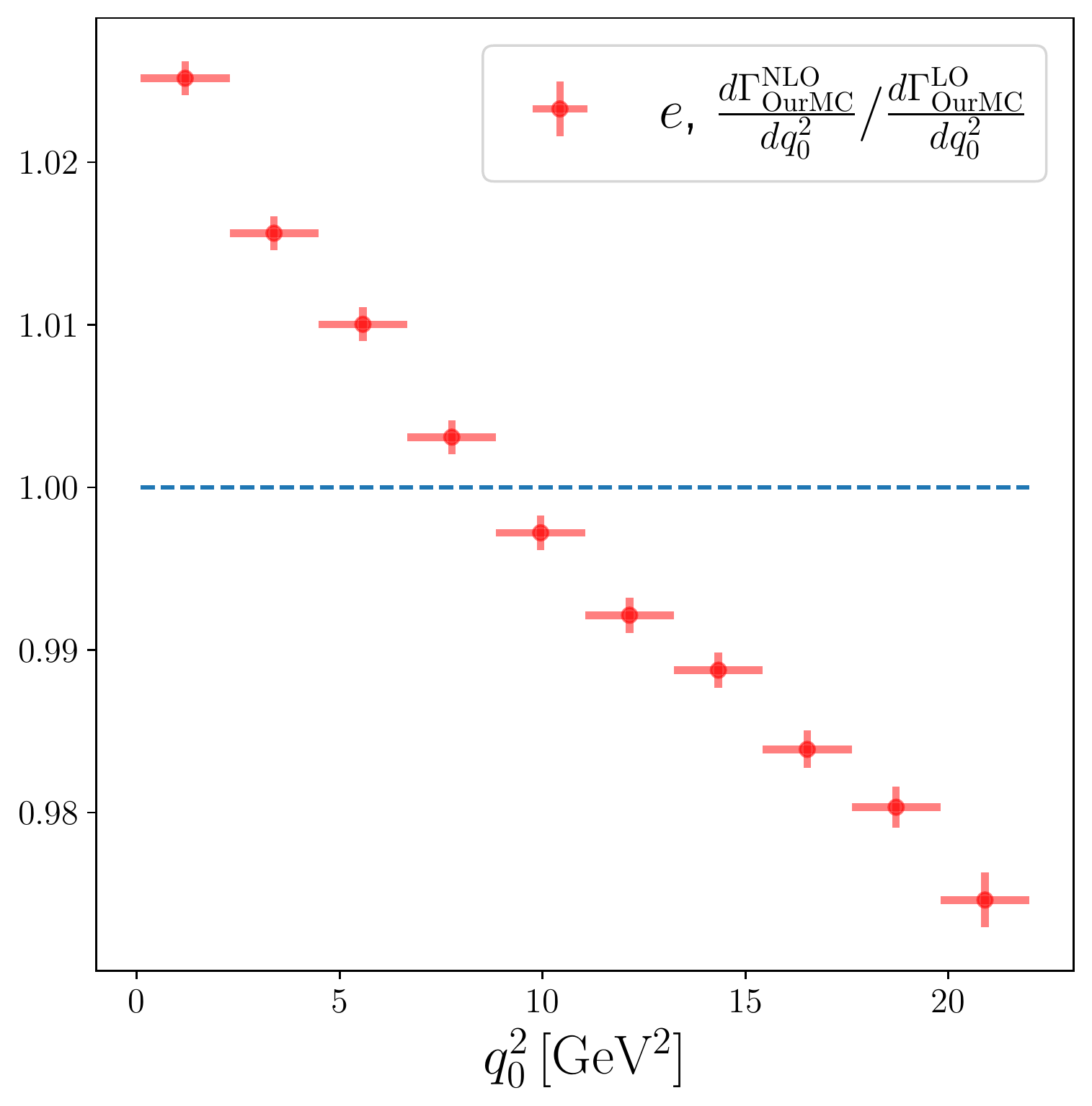}
      \put(54, 75){$\mBrec > 4.88 \GeV$}
      \end{overpic}
      \begin{overpic}[width=0.97\linewidth]{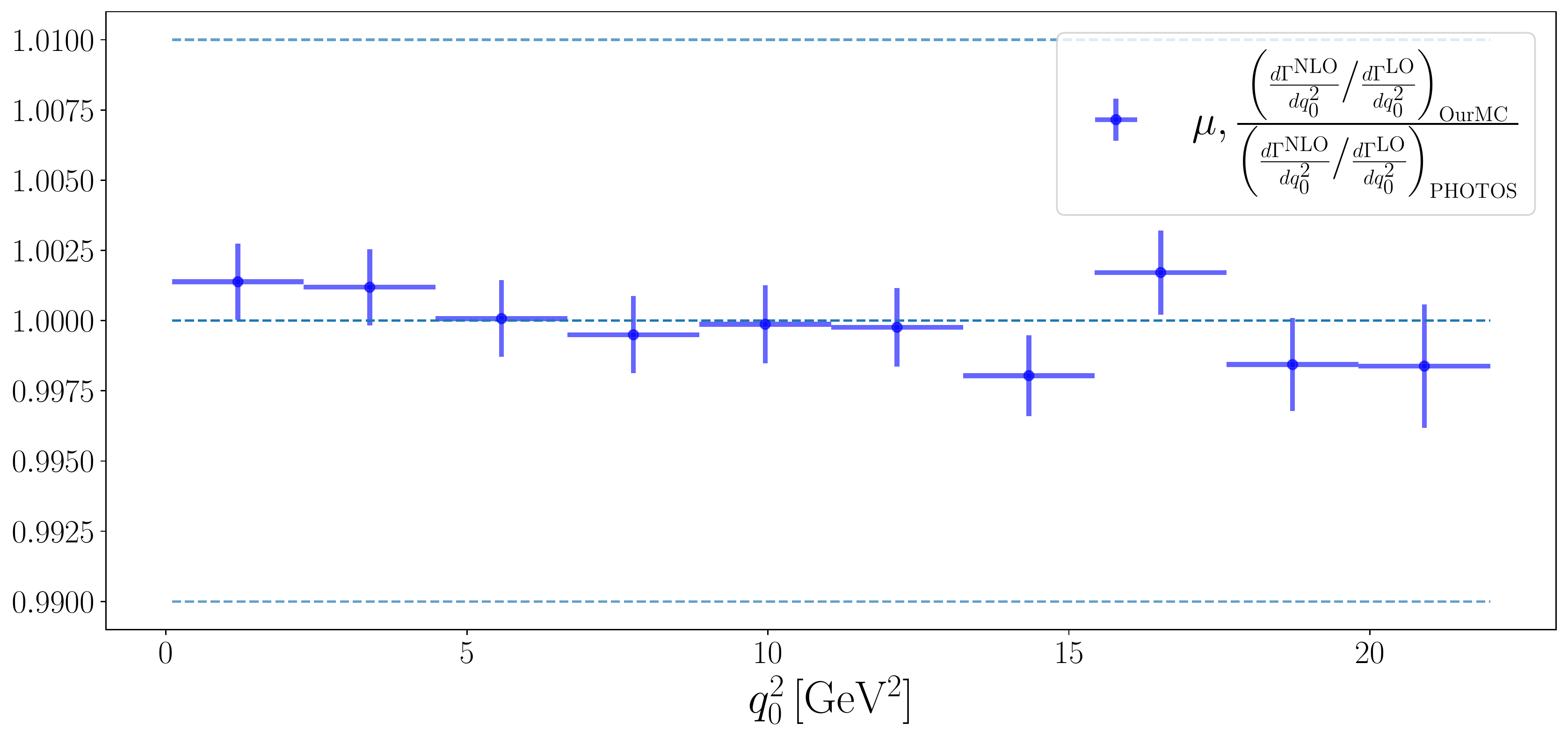}
      \put(45, 40){$\mBrec > 5.18 \GeV$}
      \end{overpic}
      \begin{overpic}[width=0.97\linewidth]{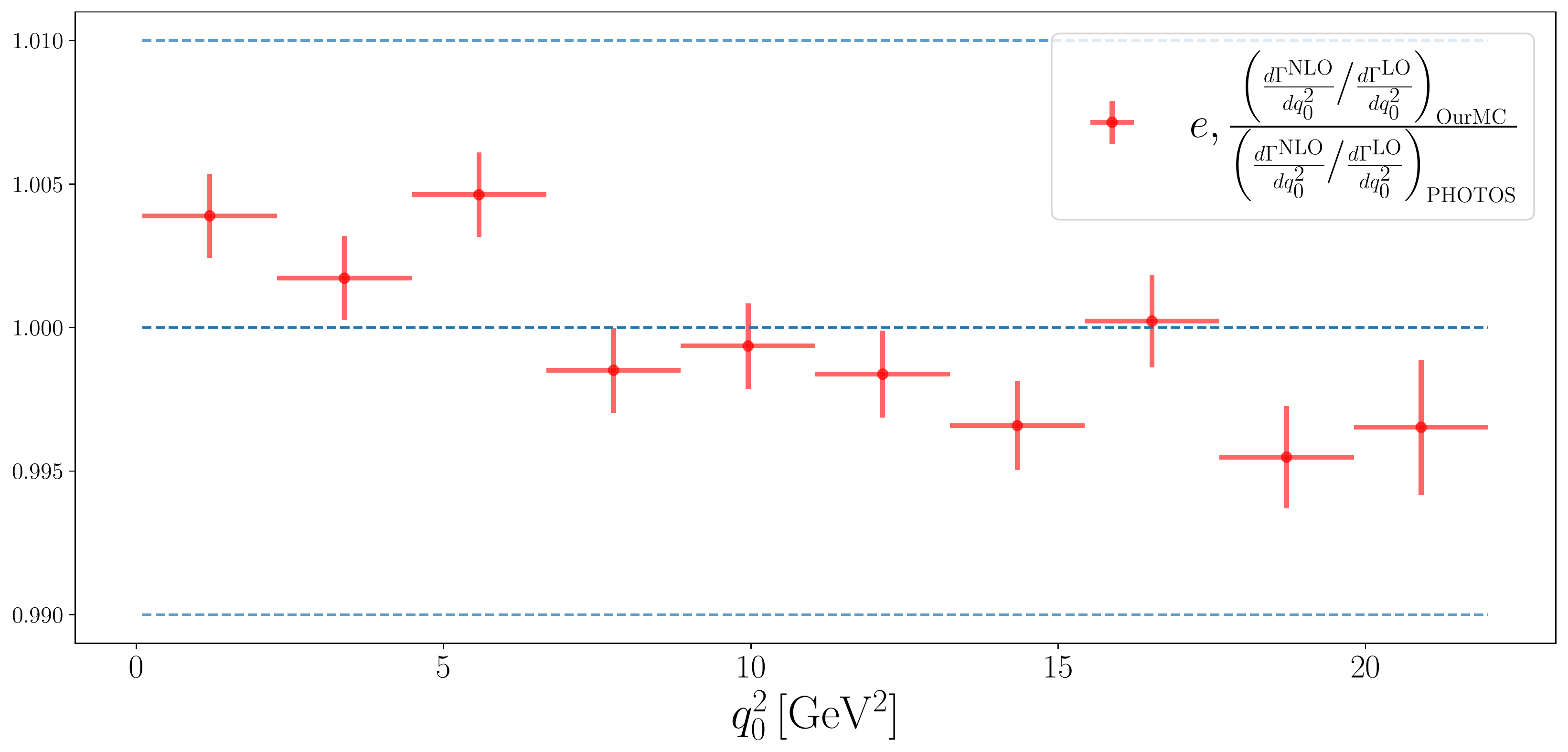}
      \put(47, 40){$\mBrec > 4.88 \GeV$}
      \end{overpic}      
	\end{center}

\caption{Differential distributions in $q_0^2$ for the short distance transition only:  
NLO over LO   for  muons in blue (top left) and  for electrons in red (top right) in our MC,
with appropriate cuts as in \TAB\ref{tab:mBrec}. 
The normalisation of these upper plots is  arbitrary (cf.~main text). 
The double ratios of our MC versus the \PHOTOS framework,  shown  in the middle and bottom plots,  are free of ambiguities.} \label{fig:q02comparison}
\end{figure}

\begin{figure}[!htbp]
   \begin{center}
      \begin{overpic}[width=0.48\linewidth]{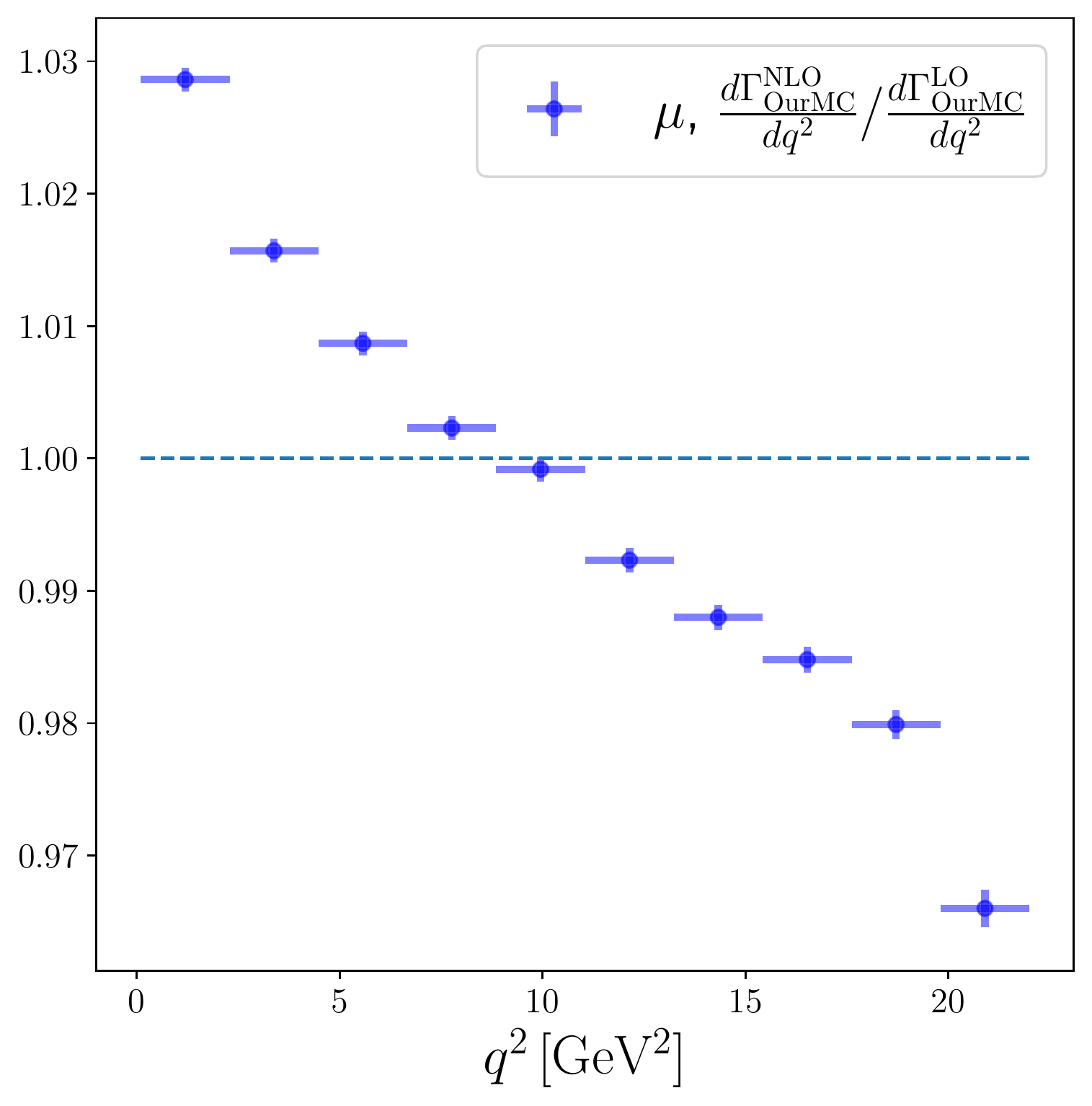}
      \put(53, 78){$\mBrec > 5.18 \GeV$}
      \end{overpic}
      \begin{overpic}[width=0.48\linewidth]{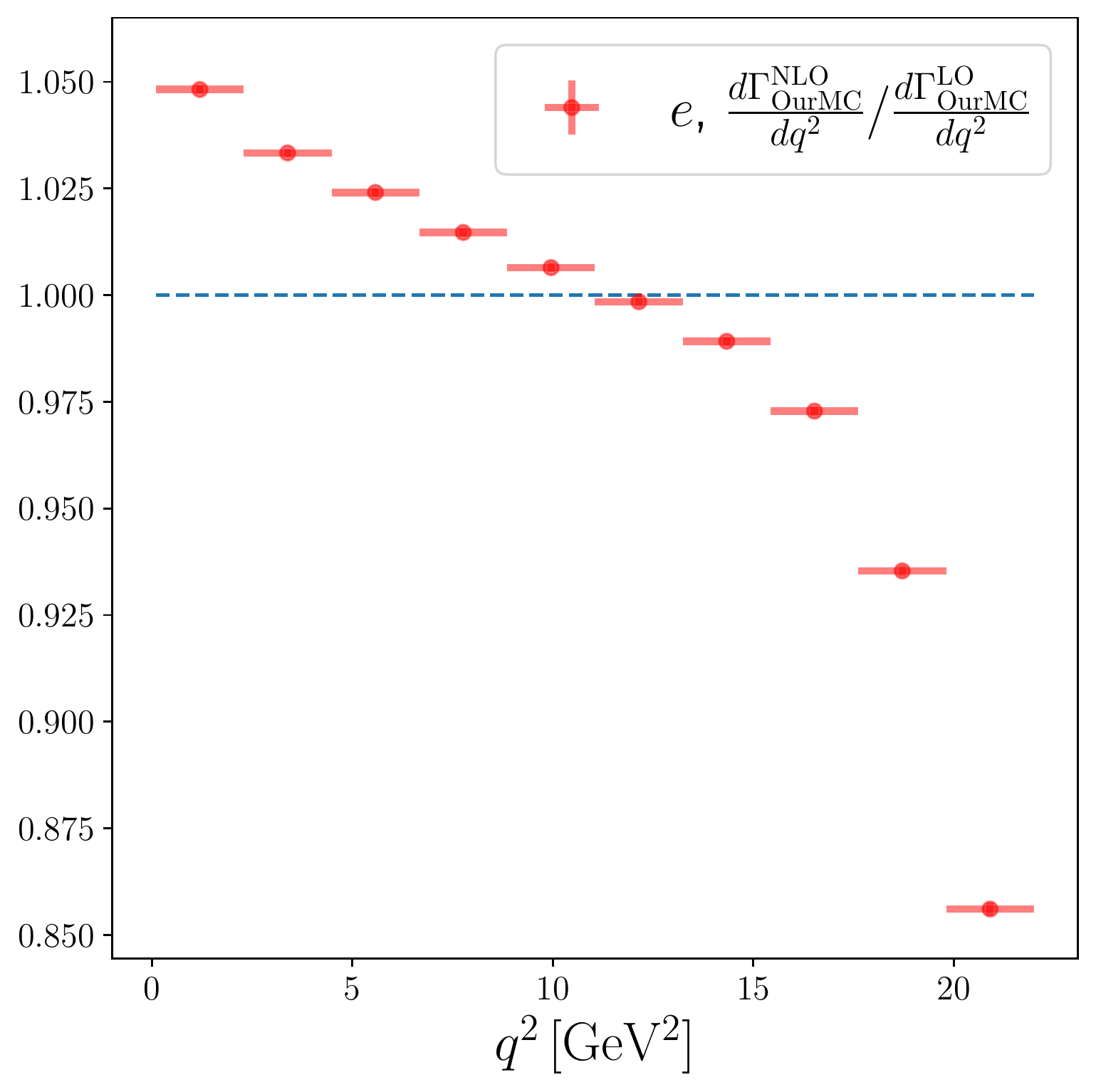}
      \put(55, 78){$\mBrec > 4.88 \GeV$}
      \end{overpic}
      \begin{overpic}[width=0.97\linewidth]{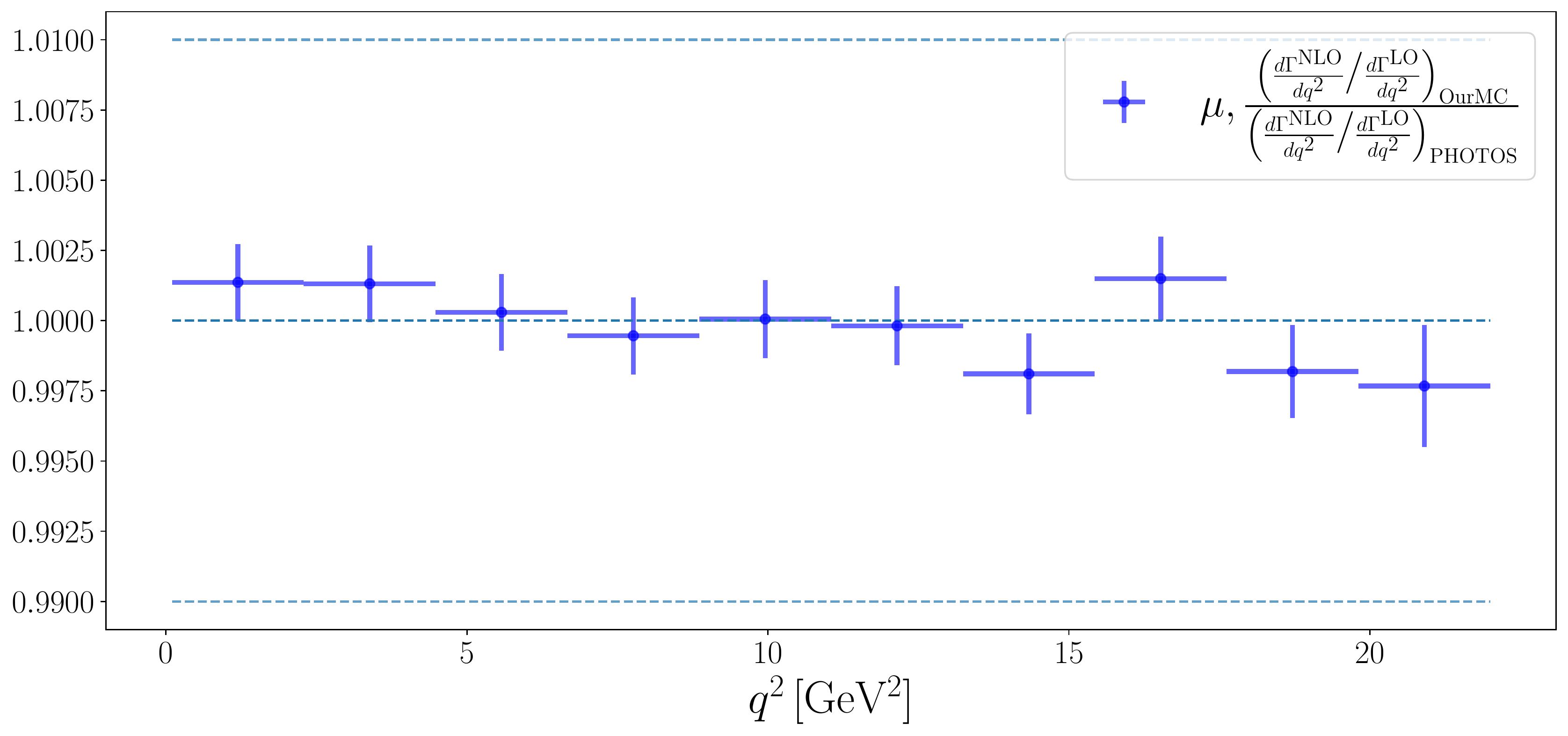}
      \put(45, 40){$\mBrec > 5.18 \GeV$}
      \end{overpic}
      \begin{overpic}[width=0.97\linewidth]{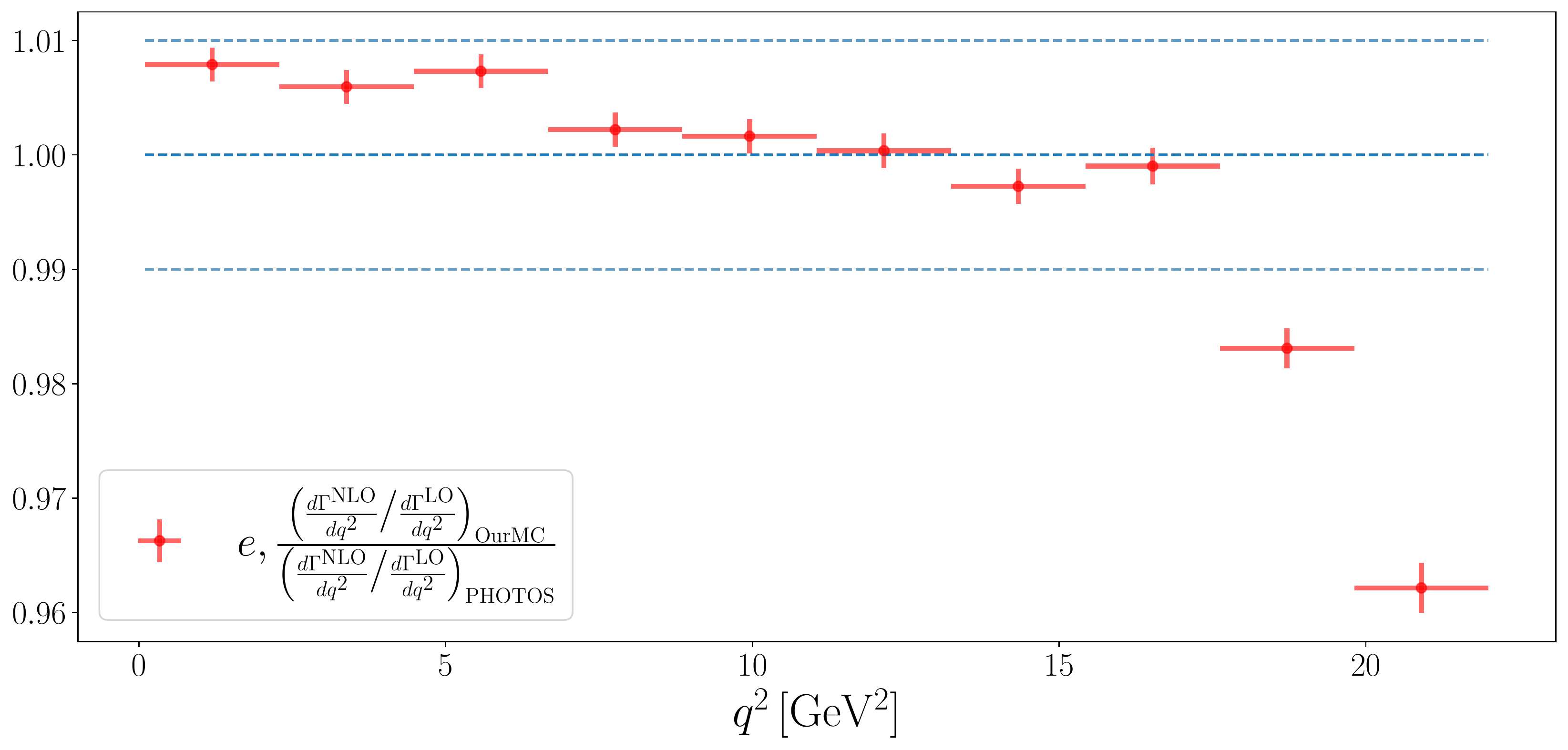}
      \put(38, 10){$\mBrec > 4.88 \GeV$}
      \end{overpic}      
	\end{center}

\caption{Differential distributions in $q^2$ for the short distance transition only: The notation is the same as in 
Fig.~\ref{fig:q02comparison}.} \label{fig:q2comparison}
\end{figure}

The main results of this section consist of  the plots in
  \FIG\ref{fig:q02comparison} and \FIG\ref{fig:q2comparison}, for the kinematic variables \qnotsq 
  \eqref{eq:q02} and \qsq \eqref{eq:q2}, respectively. 
   In each of these figures, the top plots display the impact of the radiative corrections on the \qnotsq- and \qsq-spectra when considering either muons (left) and electrons (right), in our MC. 
 The normalisation per se of the MC-plots is not meaningful as both LO and NLO are separately normalised  to 1 when integrated over \qnotsq 
 or \qsq (compare with the normalised plots in \FIG~4 in  \cite{Isidori:2020acz}).  
 This ambiguity can be removed by taking double ratios  between our MC-approach and the \PHOTOS software as shown in the middle and bottom of these figures
 for  muons and electrons respectively. 
 
 Let us discuss the $\qnotsq$-variable first. Even though the hard-collinear logs cancel in this variable,
 the introduction of a photon energy cut-off, via 
  $\mBrec = (4.88,5.18)\GeV$, leads to  sizeable QED contributions.
  The agreement between the two approaches is excellent as shown by the good compatibility of these distributions with unity across the \qnotsq-spectrum, for both lepton flavours (bottom plot).
	
The distributions in the $q^2$-variable, shown  in  \FIG\ref{fig:q2comparison}, are more delicate as hard collinear logs do not cancel. 
In addition, events can migrate in the $\qsq$-spectrum due to radiation which will be important when discussing the impact 
of the resonances. Hence, even without placing a photon-energy cut-off, the corrections are sizeable, 
cf.~\FIG~4 in \cite{Isidori:2020acz}. 

Again, the agreement between the two approaches is excellent, as expected, except for 
electrons at high \qsq where deviations up to $\ORD(4\%)$ are found. 
This originates from large corrections that go beyond the fixed $\ORD(\alpha)$
accuracy of our MC.
In fact, at the kinematic endpoint, the corrections are roughly $20\%$ (cf. 
 \FIG~8 in \cite{Isidori:2020acz}) at NLO, indicating the need of NNLO accuracy to reach $\%$ level precision, and 
explaining qualitatively the residual  difference with \PHOTOS (where the resummation of the leading-log corrections is implemented).
Using the splitting function formalism in \SEC\ref{sec:splittres}, in its resummed form, we were able to reproduce quantitatively
the $\ORD(4\%)$-effect between our MC and \PHOTOS (cf.~footnote \ref{foot:resum}).

It is worthwhile to elaborate on why  the corrections are large at the kinematic endpoint in $q^2$.
This happens because, at the  endpoint, the leptons carry all the energy and there is effectively no phase space for the real radiation. As a result, near the endpoint in $q^2$,
virtual corrections dominate and the cancellation between real and virtual corrections is maximally out of balance.
In other words, requiring large $q^2$ is equivalent to a tight cut on the photon emission energy. 
In the splitting function approach, this can be seen from the lower boundary of the real emission integral (over the momentum fraction carried by the lepton, $ z $) approaching the upper boundary.

	In summary, the  cross checks we have performed allow us to validate that the approximations adopted by \PHOTOS 
		in describing (real and virtual)  QED corrections  in \BKll decays are accurate to sub-percent level. 
	Additional plots displaying comparisons of the impact of radiative corrections on 
	the kinematic variable $\ctl$
	 between our MC and \PHOTOS can be found in \APP\ref{app:suppplots}.

\section{Adding Long Distance (Charmonium Resonances)}
\label{sec:charm}

In this section, we assess the impact of the charmonium resonances on the lower part of the spectrum, specifically on 
the $ 1.1 \GeV^2  <q^2  < 6 \GeV^2 $ region  currently used to measure the LFU ratios.
While the main contribution (peak) of the resonances is cut in the experimental analyses, a residual effect 
from the radiative tail of the resonances is potentially present
 at hadron colliders, where the $q_0^2$ variable is not accessible. 
 As previously mentioned,
the migration in $q^2$, due to QED radiation, implies that events generated at high $q^2_0$ (e.g.~close to the resonance
region) necessarily move down toward low $q^2$-values, possibly affecting the {\em signal} region for the rare mode.
The migration is controlled by the $\mBrec$ cut: 
only events with 
\begin{equation}
q^2  \leq  q_0^2  \leq   (q_0^2)_{\textrm{max}}  \equiv q^2 +  \des m_B^2~,
\label{eq:q0max}
\end{equation}
are relevant to determine 
radiative corrections at  a given $q^2$-value.\footnote{~The value $(q_0^2)_\textrm{max}$ is reached for photons 
emitted backward with respect to $\vec q$ in the $B$ RF~\cite{Isidori:2020acz}.}

\begin{table}[t]
	\centering
	\begin{tabular}{| c | ll  | l |}
	\hline
		$\ell$ \quad & ${\mBrec[\GeV]}$ & $\des$  \quad & $(q_0^2)_{\textrm{max}}$ \\		  \hline
		$ \mu$   & $5.18 $  &   $0.0486$   & $q^2 + 1.36 \,\GeV^2 $   \\
		$e$   & $4.88 $  &   $0.146$  &  $q^2 + 4.07\, \GeV^2 $\\  \hline
	\end{tabular}
	\caption{\small Relation between the cut on the reconstructed mass $\mBrec$ 
	and the maximal  value of $q_0^2$ affecting the spectrum at a given $q^2$-value, after photon radiation, according to 
	(\ref{eq:q0max}).
	The specific values of  $\mBrec$ are fixed to the same values used in the LHCb analysis of $R_K$~\cite{Aaij:2014ora}. 	}
		\label{tab:mBrec}
\end{table}

The $\mBrec$ cuts used to define the signal windows for electron and muon modes in the LHCb analyses
are reported in \TAB\ref{tab:mBrec}.  Note that the cut on the electrons is looser than the one on the muons:
a measure implemented to decrease the loss of events in the electron case where  radiation effect is stronger.  
As can easily be checked,  in the electron case, events at $q^2 = 6 \GeV^2$ 
probe (via photon radiation) the non-radiative spectrum above the $J/\Psi$-resonance 
($m_{J/\Psi}^2 \approx 9.58 \GeV^2$), but do not probe the $\Psi(2S)$ peak.

This effect is well known and the residual contribution of radiation from the $J/\Psi$ in the signal region is 
taken into account in the experimental analyses. However, this is simulated as a completely incoherent process,
while in reality interference effects between the SD amplitude and LD one are present. 
The purpose of this section is to estimate the possible impact of these effects.
In \SEC\ref{sec:charm-para}, we discuss how the amplitude of the rare mode can be adapted to 
describe SD-LD interference terms.  Using this modified parametrisation of the amplitude, in \SEC \ref{sec:MCres},
we analyse the numerical impact of the interference terms in our MC-framework.
Finally, in \SEC \ref{sec:splittres} we analyse the effect of interference terms, as well as the whole 
modulus square of the LD amplitude, via a semi-analytic approach.

\begin{table}[t]
	\centering
	\begin{tabular}{| l  | rrrr | }
	\hline
		$\Psi$ \quad & $m_{\Psi}[\MeV]$ &  ${\cal B}(\Psi \to e^+e^-)$ &  ${\cal B}(\Psi \to \mu^+\mu^-) $ &  $\Gamma_{\Psi}[\MeV]$  \\		  \hline
		$J/\Psi(1S)$   &    $3097$   &   $5.971(32)\cdot 10^{-2}$  &   $5.961(33)\cdot 10^{-2}$  & $92.6(17) $    \\
		$\Psi(2S)$   & $3686.1(6)$  &   $7.93(17) \cdot 10^{-3}$ &   $8.0(6) \cdot 10^{-3}$  &  $294(8) $\\  \hline
	\end{tabular}
	\caption{\small Data of charmonium resonances included in our analysis as taken from \cite{PDG}.
	The $J/\Psi$ mass uncertainty  is negligibly small.}
	\label{tab:mPsi}
\end{table}

\subsection{Parameterisation of the  charm amplitude}
\label{sec:charm-para}

We extend  the parameterisation of the amplitude in \SEC\ref{sec:comparison}  for the SD  form factor by including the effects of the  charmonium resonances which we label as long distance (LD)
\begin{equation}\label{eq:amplitudewithjpsi}
{\cal A}^{\textrm{tot}}_{\Bin \to \Kout\ell^+ \ell^-} = 
{\cal A}^{\textrm{SD}}  _{\Bin \to \Kout\ell^+ \ell^-} +
{\cal A}^{\textrm{LD}}_{\Bin \to \Kout  \ell^+ \ell^-}\;.
\end{equation}
The LD contribution can be absorbed into,
\begin{equation}
C^{\mathrm{eff}}_9(q^2)  = C_9 + \Delta C_9(q^2) \;,
\end{equation}
a $q^2$-dependent  Wilson coefficient  $\Delta C_9(q^2)$  (recall $C_V \propto C_9$). 
Its \qsq-dependence is parameterised  by an  $n$-times subtracted dispersion relation 
\begin{equation}
\label{eq:disp}
\Delta C_9(q^2) = \sum^{n-1}_{k \geq 0} \frac{(q^2 \mi s_0)^k}{k!} \Delta C_9^{(k)} (s_0)   + 
 \frac{(q^2 \mi s_0)^n}{ 2 \pi i }  \int_{\textrm{cut}}^{\infty}  \frac{ds}{ (s-s_0)^n }\frac{ \disc [\Delta C_9](s)}{s-q^2-i0} \;,
\end{equation}
with a cut starting just below $m^2_{J/\Psi}$. 
Above, ``disc" stands for the discontinuity,  the $k$-superscript  denotes the $k^{\textrm{th}}$ derivative, 
and ``cut" stands for the branch cut.
Formally, $n \geq 1$ as otherwise, the dispersion integral is not convergent.  
 In this form, \eqref{eq:disp} is valid in full generality and one can equally write it for the amplitude.
For   the representation \eqref{eq:disp},  the main idea is to evaluate the Taylor series in  $\Delta C_9$ for  some  $q^2 = s_0$, where perturbative methods  
can be trusted. The discontinuity,  $\disc [\Delta C_9]$, which enters the dispersion integral,  
is approximated by the  Breit-Wigner form for the resonances. This  is sufficient for the purposes 
of estimating  the contamination of the resonances on $\Bin \to \Kout \ell^+ \ell^-$ due to  QED-corrections.\footnote{~Refinements include the interference of broad resonances   \cite{LZ2014} 
and the inclusion of two-particle thresholds assuming strong constant phases 
 \cite{Cornella:2020aoq}.} The final form of $\Delta C_9$ used is the one corresponding to one subtraction
 \begin{alignat}{2}
 &  \Delta C_9(q^2)  &\;=\;& \Delta C_9(s_0)-  
  \sum_{r \in \Psi} 
\left(\frac{ q^2-s_0}{m_r^2-s_0}\right) \frac{\eta_r e^{i \de_{r}}  m_r \Gamma_{r } }{q^2 \mi m_r^2  \pl i m_r\Gamma_r}  \;. \label{eq:charmused}
 \end{alignat}
  The values of the phenomenological 
coefficients $\eta_{r}$  and $\de_{r}$ for the first two narrow resonances are reported in 
\SEC\ref{sec:MCres} and  \SEC\ref{sec:splittres} respectively (cf.~\eqref{eq:other} on how
$\eta_r$ relates to underlying parameters). The  value used  
for the subtraction term is $ \Delta C_9(0 \GeV^2) \approx 0.27 + 0.073 i$.
More details on the charm parameterisation are 
deferred to \APP\ref{app:charm}.

\subsection{Study of the $J/\Psi$-resonance interference term in our  Monte Carlo }
\label{sec:MCres}

\begin{figure}[t]
   \begin{center}
      \begin{overpic}[width=0.98\linewidth]{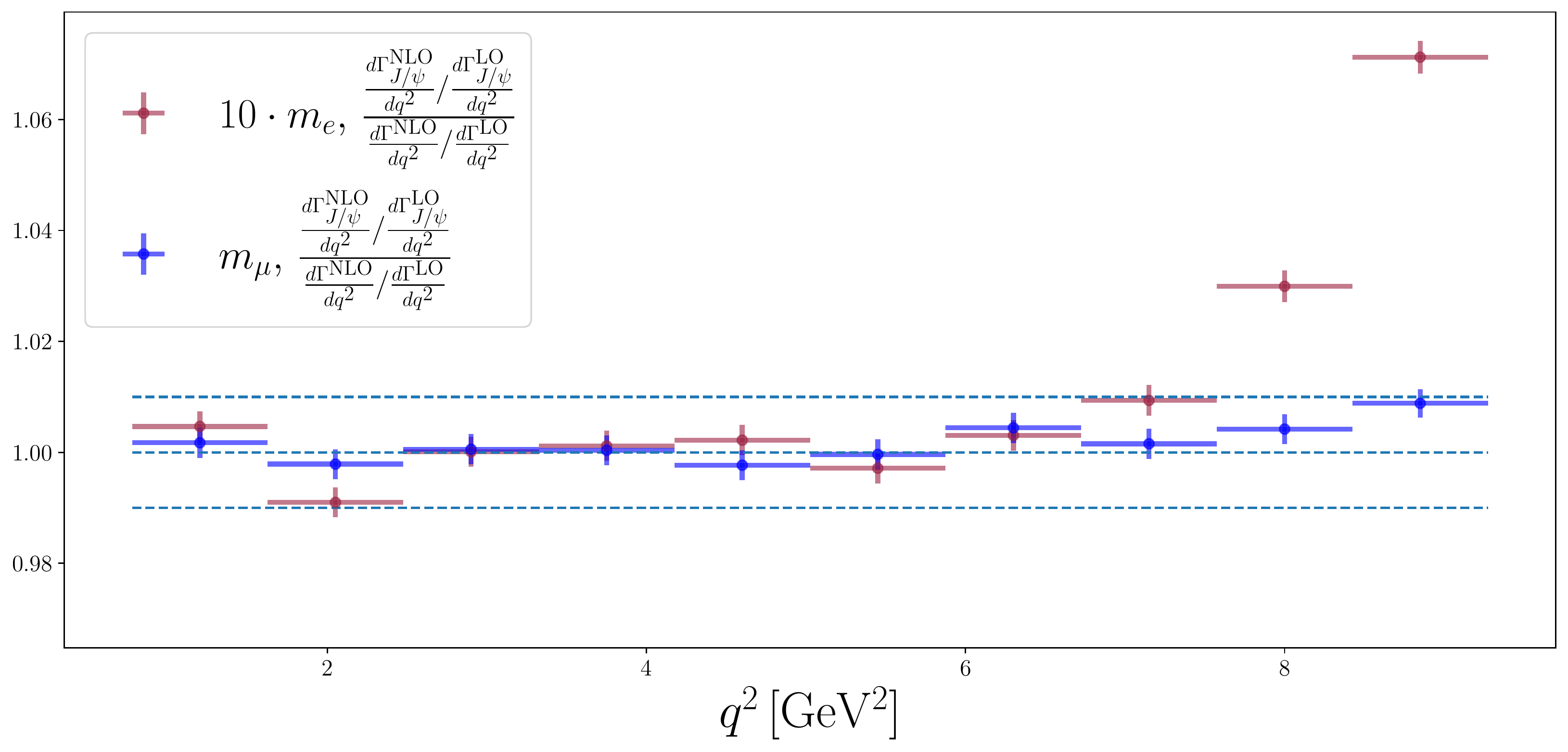}
      \put(40, 40){$m^{\mathrm{rec}}_{B}>4.88 \mathrm{GeV\,}$}
      \put(40, 35){$\delta_{J/\Psi}=1.47$}      
      \end{overpic}            
      \begin{overpic}[width=0.98\linewidth]{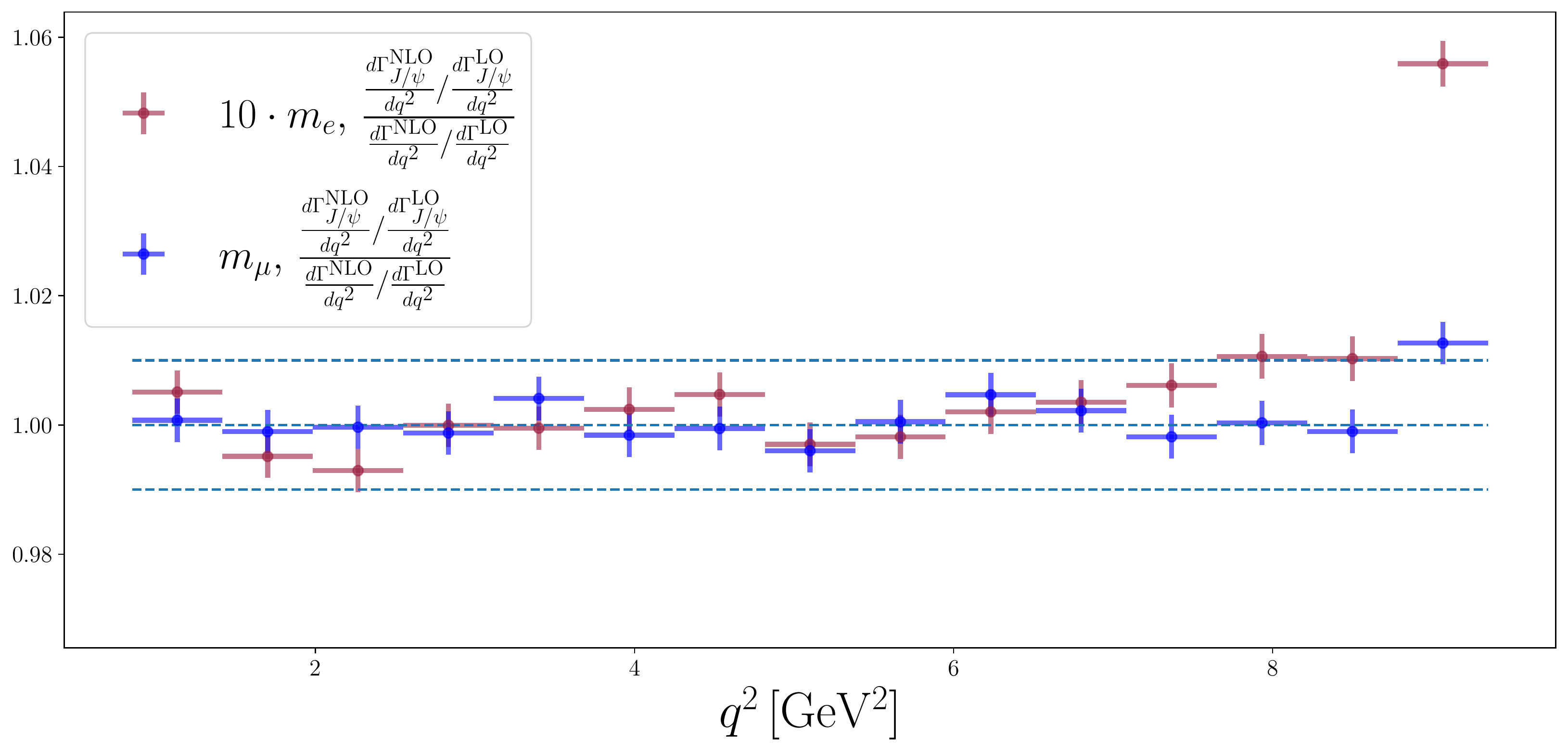}
      \put(40, 40){$m^{\mathrm{rec}}_{B}>5.18 \mathrm{GeV\,}$}
      \put(40, 35){$\delta_{J/\Psi}=1.47$}            
      \end{overpic}            
	\end{center}
\caption{Double ratio of the $q^2$-spectrum, with and without the  inclusion of interference effects induced by the $J/\Psi$-resonance,
for electron and muons, using the respective reference $\mBrec$ cuts in~\TAB\ref{tab:mBrec}. 
For numerical stability we use $m_e \to 10 m_e$ (as indicated by the darker shade in red).  The phase of the $J/\Psi$  amplitude,  relative to the SD
term, is set to $\de_{J/\Psi}=1.47$.}
\label{fig:Jpsi147}
\end{figure}

\begin{figure}[!htbp]
   \begin{center}
      \begin{overpic}[width=0.98\linewidth]{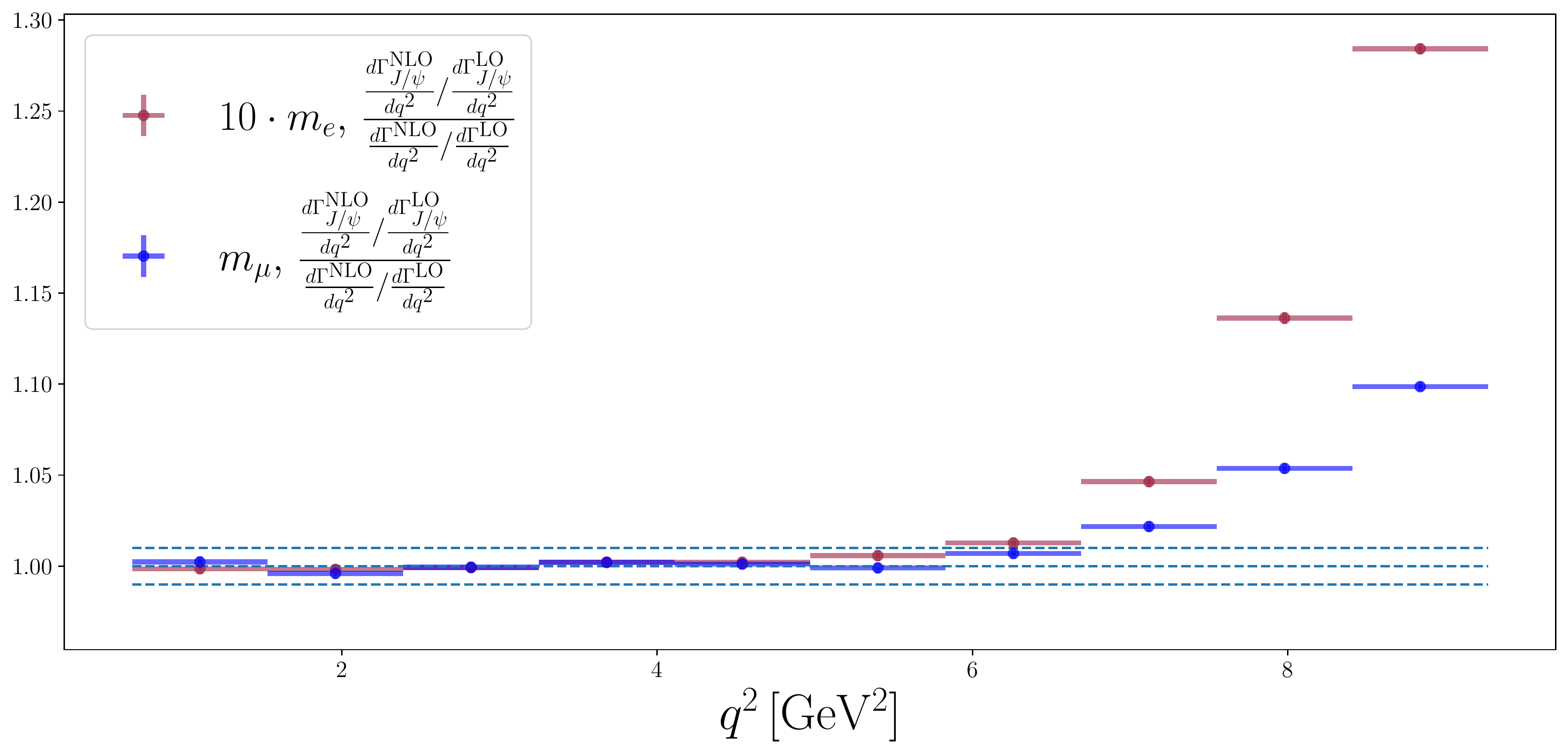}
      \put(40, 40){$m^{\mathrm{rec}}_{B}>4.88 \mathrm{GeV\,}$}
      \put(40, 35){$\delta_{J/\Psi}=0$}            
      \end{overpic}            
      \begin{overpic}[width=0.98\linewidth]{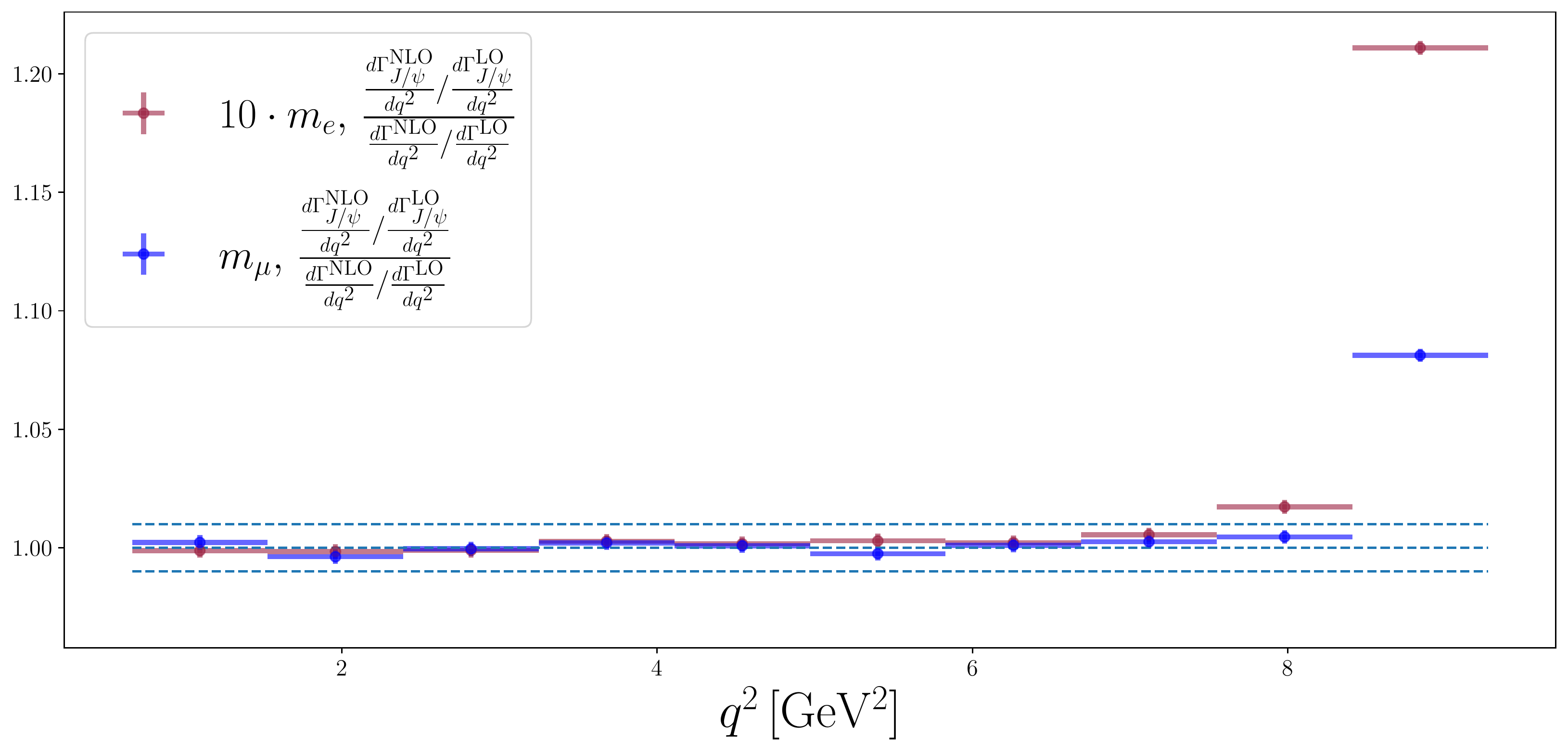}
      \put(40, 40){$m^{\mathrm{rec}}_{B}>5.18 \mathrm{GeV\,}$}
      \put(40, 35){$\delta_{J/\Psi}=0$}            
      \end{overpic}            
	\end{center}
\caption{Same plots as in~\FIG\ref{fig:Jpsi147} with the relative phase of the $J/\Psi$  amplitude set to $\de_{J/\Psi}=0$.
} \label{fig:Jpsi0}
\end{figure}

In the MC-study, we include the $J/\Psi$-resonance  in the sampling method  outlined  in \SEC\ref{sec:numprocedure}, by extending the definition of $C^{\mathrm{eff}}_9(q^2)$ as detailed in the previous section. 
The modulus squared of the resonant mode, 
${\cal A}_{\Bin \to \Kout (J/\Psi\to \ell^+ \ell^-)} \subset {\cal A}^{\textrm{LD}}_{\Bin \to \Kout  \ell^+ \ell^-} $ in \EQ\ref{eq:amplitudewithjpsi}, is not included in our simulation since 
 its sharp pole at $q^2 \approx m^{2}_{J/\Psi}$ renders the MC sampling efficiency too low.  
In turn, this requires to place a cut on $\qnotsq \leq 9.59\GeV^2$ (cf.~\APP\ref{app:fthtable} for details) since the remaining terms become negative above that threshold,  invalidating their interpretation as a PDF (cf.~\SEC\ref{sec:numprocedure}). Another factor limiting the sampling efficiency is the lepton mass, which in the electron case has to be increased to 10 times its physical value (c.f.  \APP\ref{app:fthtable}) to allow for efficient sampling.\footnote{~ 
We have checked that the  $\ln m_\ell$-behaviour 
is consistent with what is obtained using the the semi-analytic method  described in \SEC\ref{sec:splittres}.}

Our approach is well justified since the modulus squared of the $J/\Psi$-resonance is  well simulated by \PHOTOS
(see \cite{LHCb:2021trn}), and the component describing its leakage in the signal region is included in the fit used to extract the rare mode yield. With our simulation, we aim to analyse the effect on the \qsq bin migration of the $J/\Psi$ interference term, that has so far  not been considered in the experimental analyses.

 The resonance data is given by the normalisation  $\eta_{J/\Psi} = 8180$  
  in the  notation of \eqref{eq:charmused} 
  (or  $\rho_{J/\Psi}= 1.38$,   in the notation of \eqref{eq:other}), with
mass and width in~\TAB\ref{tab:mPsi},
 and the  interference phase $\de_{J/\Psi}$.   For the latter, we choose two representative values:  
 $\de_{J/\Psi} = (1.47,0)$  where the former is  deduced by the LHCb analysis of the dilepton spectrum~\cite{LHCb:2016due}, 
 and the latter is a conservative choice aimed at maximising the $J/\Psi$ interference effect. 
 
The plots in \FIGs\ref{fig:Jpsi147} and \ref{fig:Jpsi0} show the effect, as a function of \qsq, on the radiative corrections when including the $J/\Psi$ interference term in the decay width, for interference phases of $\de_{J/\Psi} = (1.47,0)$ respectively. More specifically, they represent the double ratio of NLO over LO differential decay widths including charm over the same ratio without charm.

Both figures include electron-like (red) and muon (blue) distributions, 
with appropriate  $\mBrec = (4.88,5.18)\GeV$ cuts (cf.~\TAB\ref{tab:mBrec}). As can be seen from these plots, the impact of the SD--LD interference term is well below the
$1\%$-level in the  $q^2 < 6 \GeV^2$ region, for the (realistic) phase choice $\de_{J/\Psi} = 1.47$.
Even, in the  conservative case $\de_{J/\Psi} = 0$, it remains just below $1\%$.
We thus conclude that when applying the aforementioned cuts, the experimental approach 
of neglecting the \emph{interference} effect of charmonium resonances, 
when fitting for the rare mode  in the  $q^2 < 6 \GeV^2$ region, is well-justified.

\subsection{$J/\Psi$ and $\Psi(2S)$, including the  resonant  mode via a semi-analytic approach}
\label{sec:splittres}

Here, we follow a semi-analytic approach, using the splitting function, 
which reproduces the relevant collinear logarithms.  
There are a few advantages to  this approach: it is  numerically less demanding,  
there is no normalisation ambiguity,  no issues with positivity, 
further  resonances are  easily incorporated, we can simulate for the actual electron mass 
and  we may assess the impact of the 
full resonant amplitude.
While it is not a MC-based approach, 
and therefore not directly of use for  an  event simulation, it  may be serve as a reweighting tool for our MC-framework.

Evaluating the impact of the full resonant amplitude is of interest  since its rate is sizeable  ${\cal B}(\bar B^0 \to {{\bar K}^{0}}{ J / \psi})=  8.91(21) \cdot 10^{-4}$ compared to the rare mode itself which is $\ORD(10^{-6})$,
and with a cut set at  $\mBrec = 4.88\GeV$  
the electron mode does probe the first resonant peak for  $\qsq \approx 6 \GeV^2$,  as previously stated.  
In fact, in a decay like $\bar B \to \bar K \ell^+\ell^-$ the (hard)-collinear and soft-collinear logs in the lepton mass can be reproduced from the lepton to lepton-photon 
splitting function.\footnote{~The specific details are
postponed to a future publication \cite{NZ22} and  for more generic remarks we refer the reader to \cite{Zwicky:2021olr}. 
Although, note that the kinematic relations, to follow below, can be found in our previous work \cite{Isidori:2020acz} (cf. ancillary 
notebook for the expression with $m_K \neq 0$).
\EQ(A.5)  in \cite{Isidori:2020acz} corresponds to the single-differential and photon-inclusive version of \eqref{eq:magic}.}$^,$\footnote{~\label{foot:resum}  This formalism can be extended to resum all the collinear logs using the electron structure function.
Taking  the last bin $[20.9 \GeV^2,(m_B \mi m_K)^2]$, used in \FIG\ref{fig:q2comparison}, and weighing by the rate, we produce an effect of $\approx 0.96$ which agrees very well with the central 
value in that figure.}
 
It is convenient to parameterise the relative QED correction 
$d^2 \Gamma \propto (1 + \Delta^{(\TT)}(\hat{q}^2,\cl) ) d \hat{q}^2 d \cl)$, following our earlier work, as
\begin{equation}
\Delta^{(\TT)}_{\textrm{hc}}(\hat{q}^2,\cl)  = 
 \frac{ \al }{\pi \,}
\left( \frac{1}{  \Gamma^{\LO }} \frac{d^2 \Gamma^{\LO}(\hat{q}^2) }{d\hat{q} d \cl ^2}  \right)^{-1} 
\big( \hat{Q}_{\lone}^2 \tilde{\Delta}^{(\TT)}_{\textrm{hc},\lone}
+ \hat{Q}_{\ltwo}^2 \tilde{\Delta}^{(\TT)}_{\textrm{hc},\ltwo}  )  \,.
\end{equation}
where $\hat q^2 \equiv q^2/m_B^2$ for brevity and the subscript ``hc" stands for the 
(hard) collinear contribution. This quantity reads 
\begin{equation}
\label{eq:magic}
 \tilde{\Delta}^{(\TT)}_{\textrm{hc},\,\lone}(\hat{q}^2,\cl)  =
 \ln \frac{\muhc}{\mlone}   \left(\frac{1}{  \Gamma^{\LO}} \int_{\max(\hat q^2,  z_{\lone}^\de)  }^{1} dz  P_{f \to f \ga}(z)   \frac{d^2 \Gamma^{\LO}(\qzh^2,\clz) }{d \qzh^2 d\clz } \right) J_{\lone}(\cl,z)   \;,  
\end{equation}
where  
$ P_{f \to f \ga}(z)  \;=\;  \lim_{z^* \to 0} \left[  \frac{1 + z^2}{(1- z)}\theta((1-z^*)-z)  +( \frac{3}{2} + 2 \ln z^*)\de(1-z)   \right]$ is the splitting function where 
 $\muhc = \ORD(m_B)$ is an a priori undetermined scale (to be commented further below). 
The variable  relations and the  Jacobian ($d \qz^2 d \clz \;=\; J_{\lone} (\cl,z) dq^2 d \cl$) are given by
\begin{equation}
\label{eq:varJac}
q^2 = z q_0^2 \;, \quad  \clz|_{m_K=0} =  \frac{\cl(1+z) + \bar z}{\cl  \bar z + 1+z} \;, \quad 
J_{\lone}(\cl,z)|_{m_K=0}   =  \frac{4}{( \cl \bar z  + 1+z)^2} \;,
\end{equation}
with $\bar z \equiv 1-z$. The lower integration boundary is set  by the maximum of 
the   photon inclusive limit $\hat{q}^2$  and the photon-cut off dependent\begin{equation}
\label{eq:z1}
z_{\lone}^\de|_{m_K=0}  = 
\frac{1+\hat{q}^2-\de+\cl (1-\hat{q}^2-\de)}{1+\hat{q}^2+\de+\cl (1-\hat{q}^2-\de)} \;.
\end{equation}
The corresponding expression for  $\tilde{\Delta}^{(\TT)}_{\textrm{hc},\,\ltwo}$ can be obtained 
by changing the signs on all the cosines in the lepton angles in \EQs\eqref{eq:varJac} and \eqref{eq:z1}.

\begin{figure}[!htbp]
   \begin{center}
    \hspace{-7.8cm}   \begin{overpic}[width=0.5\linewidth]{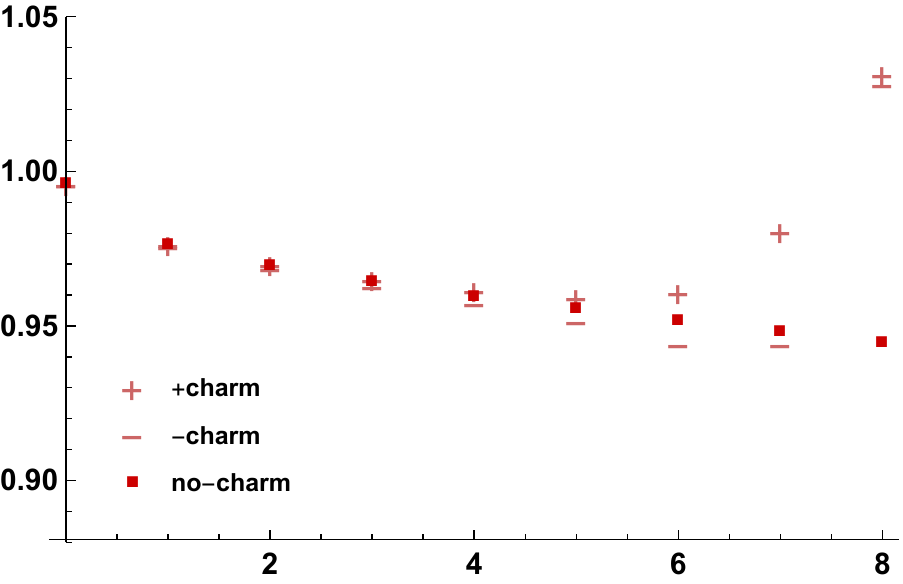} 
      \put(12,51){${ \small \boxed{e, \frac{\frac{d\Gamma^\textrm{NLO}}{dq^2}}{\frac{d\Gamma^{LO}}{dq^2}}}}$}
     \put(43, 57){${\scriptstyle{ \mBrec >4.88 \text{GeV} }  }$}
       \put(43, 51){${ \scriptstyle{ \delta_{J/\Psi,\Psi(2S)}=0,\pi}}$}  
       \put(43, 45){$ \scriptstyle{\text{cut on } |A_{\bar B \to \bar K (\Psi \to ee)} |^2 } $ }
       \put(77,-5){${ \small q^2[\GeV^2] }$}
      	\hspace{7.8cm}
	\begin{overpic}[width=0.5\linewidth]{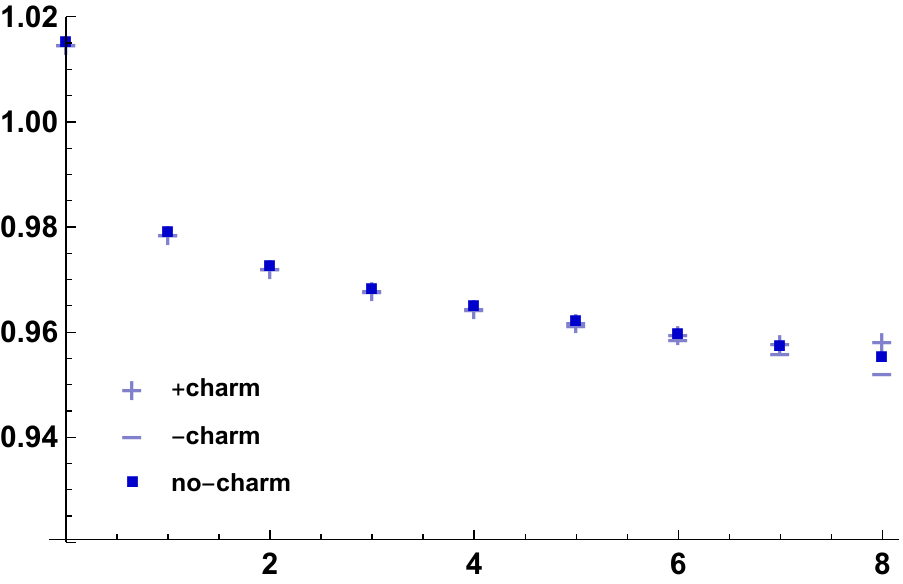} 
     \put(12,51){${ \small \boxed{\mu, \frac{\frac{d\Gamma^\textrm{NLO}}{dq^2}}{\frac{d\Gamma^{LO}}{dq^2}}}}$}
     \put(43, 57){${\scriptstyle{ \mBrec >5.18 \text{GeV} }  }$}
       \put(43, 51){${ \scriptstyle{ \delta_{J/\Psi,\Psi(2S)}=0,\pi}}$}  
       \put(43, 45){$ \scriptstyle{\text{cut on } |A_{\bar B \to \bar K (\Psi \to \mu\mu) }} |^2$}
       \put(77,-5){${ \small q^2[\GeV^2]  }$}
   	\end{overpic}       
      \end{overpic}            
	\end{center}   
\caption{\small  Plots   with the resonant mode cut out (cf. main text for explanation).
 For $q^2 < 6 \GeV^2$ the interference effects are  small, even in the electron case 
 (confirming the plot in \FIG\ref{fig:Jpsi0}), and do not indicate any contamination to $R_K$ in particular. 
 The corresponding plot without the LO normalisation can be found in 
 \APP\ref{app:suppplots} in \FIG\ref{fig:sa_window_abs}.}
	\label{fig:sa_window}
\end{figure}

 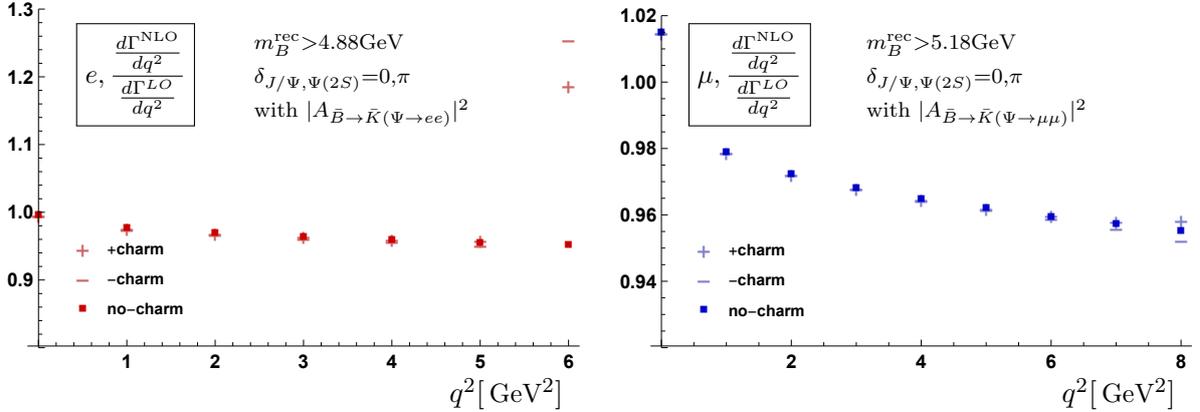
\begin{figure}[!htbp]
   \begin{center}
    \hspace{-7.8cm}   \begin{overpic}[width=0.5\linewidth]{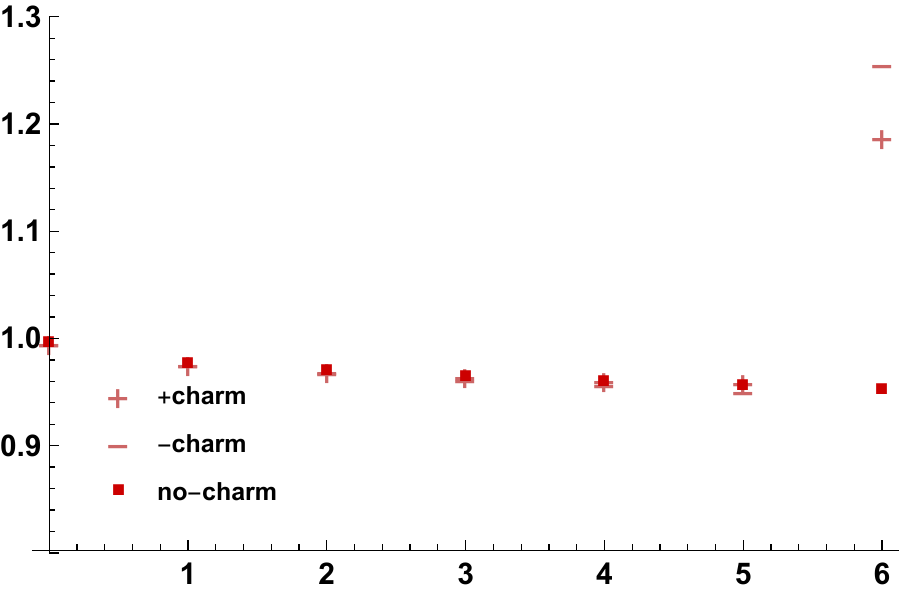} 
     { \put(12,51){${ \small \boxed{e, \frac{\frac{d\Gamma^\textrm{NLO}}{dq^2}}{\frac{d\Gamma^{LO}}{dq^2}}}}$}
       \put(43, 57){${\scriptstyle{ \mBrec >4.88 \text{GeV} }  }$}
      \put(43, 51){${ \scriptstyle{ \delta_{J/\Psi,\Psi(2S)}=0,\pi}}$}  
       \put(43, 45){$ \scriptstyle{\text{with }|A_{\bar B \to \bar K (\Psi \to ee)} |^2}$ }  
       \put(77,-5){${ \small q^2[\GeV^2] } $} }
      	\hspace{7.8cm}	\begin{overpic}[width=0.5\linewidth]{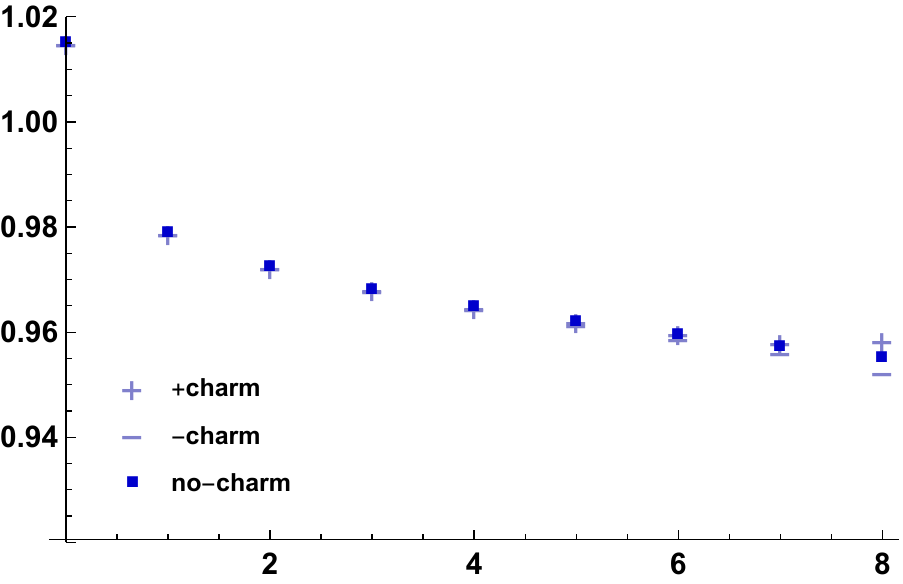} 
   {  \put(12,51){${ \small \boxed{\mu, \frac{\frac{d\Gamma^\textrm{NLO}}{dq^2}}{\frac{d\Gamma^{LO}}{dq^2}}}}$}
      \put(43, 57){${\scriptstyle{ \mBrec >5.18 \text{GeV} }  }$}
       \put(43, 51){${ \scriptstyle{ \delta_{J/\Psi,\Psi(2S)}=0,\pi}}$}  
       \put(43, 45){$ \scriptstyle{\text{with }|A_{\bar B \to \bar K (\Psi \to \mu\mu)} |^2} $}  
       \put(77,-5){${ \small q^2[\GeV^2] } $} 
       }
   	\end{overpic}       
      \end{overpic}            
	\end{center}
\caption{\small  Same plots as in \FIG\ref{fig:sa_window} including resonant modes:
  (left) for electrons and (right) for muons  respectively. It is noted that at $q^2 = 6 \GeV^2$,
  the effect is noticeable for electrons and care has to be taken (cf.~main text). 
  For the electrons, the plot ends at $q^2 = 6\GeV^2$ since beyond this value 
  the effects are too large (at $q^2 = (7,8)\GeV^2$ we find, approximately,  $ (6.5,40)$ and $( 8.4,73)$ 
  for +charm and -charm  respectively).  We have checked that resummation slightly tames the effect but qualitatively, it remains the same.
  The muon plot looks deceptively similar to the one in  \FIG\ref{fig:sa_window}  for $q^2 < 8 \GeV^2$ but 
  differences arise thereafter.  The corresponding plot without the LO normalisation can be found in 
 \APP\ref{app:suppplots} in \FIG\ref{fig:sa_nowindow_abs}.
	}
	\label{fig:sa_nowindow}

\end{figure}

We turn to the practical implementation. As compared to the previous section, 
we include the second resonance $\Psi(2S)$, cf.~\TAB\ref{tab:mPsi} for the basic inputs. In this case 
 $\eta_{\Psi(2S)} = 1160$ (or  $\rho_{\Psi(2S)} = 1.56$, in the notation of \eqref{eq:other})  
 describes its residue up the free phase $\de_{\Psi(2S)}$.
In order to emulate the LHCb procedure in eliminating the $J/\Psi$ mode, we cut out the amplitude squared of
the resonant mode 
in the following $\qnotsq$-window: $m_\Psi ^2- \cutter < q_0^2 < m_\Psi ^2 + \cutter    $ with  $\cutter = 0.1 \GeV^2$.
Empirically, we find that choosing the undetermined scale to be
 $\muhc^2 \approx 6 q^2$ ,does reproduce our short distance results in \cite{Isidori:2020acz} rather well. 
 It is not surprising that the scale is proportional to $q^2$ 
since this is the relevant scale ``seen" by the lepton pair. 

In order to assess the possible uncertainty of the charm contribution, we 
choose the phases to give rise to maximal interference i.e. $\de_{J/\Psi,\Psi(2S)} =0$ 
and plot the three graphs: one without any charm, one as described above and one with the sign reversed 
(i.e.   $\de_{J/\Psi,\Psi(2S)} =\pi$).  The maximal difference can then be seen as a conservative estimate for the error of 
not including the charm. 

Plots for the electron and the muon cases with the resonances cut out  are shown in \FIG\ref{fig:sa_window}.
This situation mimics the interference of the rare and resonant mode
and it is seen from the plots that, for  $q^2 < 6 \GeV^2$,  this contribution is small. 
At $q^2 = 6\GeV^2$ the difference between the two charm contributions with opposite sign is $2\%$  and when this effect is averaged over the entire $[1.1,6]\GeV^2$ bin it is clear that the effect does not exceed  $\ORD(1\%)$ which would be comparable to structure dependent corrections.  This is fortunate since, as previously mentioned, resonant versus rare-mode interference  are not included in the  LHCb analysis.  These results can be seen as a validation of the double ratio plots in \FIG\ref{fig:Jpsi0} obtained in the MC-framework (with meaningful absolute normalisation).

We turn to the case where we include the full resonant mode. 
Crucially,  the squared resonant amplitude is independent of the $\de_{J/\Psi}$-phase 
and dominates over  the interference. This can be seen from the electron plot  shown in \FIG\ref{fig:sa_nowindow}, by comparing it to the corresponding one in \FIG\ref{fig:sa_window}. 
Furthermore, it can be seen that for the electron cut-off, QED effects begin to be sizeable below the $q^2 < 6 \GeV^2$ and thus care has to be taken.  
The effect is coming from the  $J/\Psi$ resonance and the effect 
of the $\Psi(2S)$ resonance is moderate for the given electron cut.

This is further reflected in the LHCb mass fit projections to the signal mode (cf.~\FIG 2 in \cite{LHCb:2021trn}) where the leakage of the resonant mode is included in the total fit model, together with the other backgrounds components, to extract the electron signal yield.  
Amongst the  cross-checks  performed in the measurement of $R_K$ are the integrated and differential ratios $r_{\footnotesize{\JPsi}}$, which directly compare electron and muon detection efficiencies, and thus constitute a stringent validation of their analysis. The value of $r_{\footnotesize{\JPsi}}$ is known in the SM to be unity to a very high degree of accuracy, since it originates from the tree level mediated resonant mode, and was measured in the $R_K$ analysis, $r_{\footnotesize{\JPsi}} = 0.981 \pm 0.020$. This result is one sigma compatible the SM prediction and with the previous measurements of this quantity \cite{PDG}.  Moreover, as can be seen from \FIG 9-10 of \cite{LHCb:2021trn}, $r_{\footnotesize{\JPsi}}$ is also performed differentially as a function of variables which are used in the determination of the dilepton invariant mass, such as the opening angle of the lepton pair and their transverse momentum. The flatness of $r_{\footnotesize{\JPsi}}$ in those variables reflects an excellent description of the efficiency-related effects in muons and electrons. Not only are these crucial cross-checks per se, but they also validate the double ratio method used to minimise the efficiency-related systematic error in the $R_K$ measurement \cite{LHCb:2021trn}, and the accuracy in the description of the QED corrections from the absolute square of the $J/\Psi$-mode. 

Despite all these positive cross-checks, since the overall impact of the resonance modes is large (cf.~\FIG\ref{fig:sa_window}), 
it would be of great relevance  if the LHCb collaboration could perform a $q^2$-binned analysis of $R_K$. 
This would provide a further important test of the robustness of $R_K$. 

Related to that, we have investigated the robustness of the results with respect 
to non-perturbative aspects which are difficult to control:  the extrapolation of the Breit-Wigner form
and  neglecting higher resonances.  
 The Breit-Wigner resonance gives a good approximation 
close to its pole only and its extension away from the pole is not thoroughly known.  
When using 
\eqref{eq:charmused} an implicit assumption on its form was  made;
it is dictated by the short distance form factors  (cf. \APP\ref{app:charm} for further comments).  
We may therefore assess the effect by choosing flat form factors multiplying the Breit-Wigner resonances and adjusting the residue  to reproduce the $B \to \Psi e^+e^-$ rate.
Specifically, we replace the form factor by its value at the subtraction point. 
It is found that this effect leads 
to changes which do not exceed the $2\%$-level for $q^2 < 6\GeV^2$ and is thus fortunately moderate. 
The effect of changing the number of subtractions   in \eqref{eq:disp} 
can be seen as a way to estimate neglecting higher resonances since  
many of them are needed  to reproduce the 
precise asymptotics of perturbative QCD.   
However,  one versus no subtraction leads to  small $\ORD(1\%)$ changes only 
and might be seen as an indication of the consistency of the subtraction value with the resonance data. 
In summary the extension of the $J/\Psi$-resonance has a much larger effect than neglecting higher states.

  All in all, this underlines the importance of a refined $q^2$-binning  from a different viewpoint. 
Another way to look at it is that it emphasises the importance of knowing the LO amplitude 
(i.e.~the idealised  amplitude in the absence of QED)
since its precise form affects  the detailed form (size and magnitude) of QED corrections. 
In this respect, given the smaller impact of QED effects and the better experimental resolution,
the muon case can serve as a tool for a precise determination of the LO spectrum  in a data-driven approach.

 \section{Outlook and Conclusions  }
\label{sec:conclusions}

In this article, we investigated 
numerical aspects of QED corrections on the   $\BKll$ decay,  which is of particular relevance 
in view of LFU tests.
We constructed a dedicated Monte Carlo framework based on the computation in \cite{Isidori:2020acz} 
(cf.~\SEC  \ref{sec:MCframe}), and further analysed QED effects by means of a  
semi-analytic  
(splitting-function based) approach (cf.~\SEC \ref{sec:splittres}) which captures the 
 numerically dominant  collinear logs.

In \SEC\ref{sec:comparison}, we compared our Monte Carlo framework with   \PHOTOS  
at the level of the short distance matrix element (rare mode) and found good agreement at the differential level 
in all relevant variables (in particular $q^2, c_\ell$ and $q_0^2,\clz$). Particularly relevant is the comparison in the $q^2$-distribution, illustrated 
in~\FIG~\ref{fig:q2comparison}, which plays a key role in the LFU tests at hadron colliders.
Since \PHOTOS and  our approach are supposed to capture all the leading logs, agreement was to be expected. 
Indeed, a partial cross-check of \PHOTOS, in the  $q_0^2$- and with an effective cut-off in the $q^2$-distribution were already reported in~\cite{BIP16}.  
Our double differential comparison thus provides a solid   cross-validation of both our  Monte Carlo framework and \PHOTOS. 

In addition to the short distance contribution, our Monte Carlo and semi-analytic framework has allowed us to assess the impact of the resonant mode 
$\Bin \to \Kout (J/\Psi\to \ell^+ \ell^-)$ on the extraction of  $R_K$ (in \SECs
\ref{sec:MCres} and   \ref{sec:splittres} respectively).
Both the Monte Carlo and the semi-analytic approach confirm that the interference effects between resonant and rare mode 
(not included in \PHOTOS) are below $1\%$ for $q^2 < 6 \GeV^2$. This justifies not simulating these effects in the experimental setup, as  currently done. 
On the other hand, the resonant mode, as known and expected,  has a significant effect in the electron mode  below $6 \GeV^2$
as the plot in \FIG \ref{fig:sa_nowindow} quantifies.  
This is taken care of in the present experimental analyses.
As  pointed out, a useful validation of this procedure could be obtained with 
an extraction of $R_K$ in different $q^2$-bins (in the $q^2 < 6 \GeV^2$ region).  
A further independent  cross-check could be obtained  
  varying $\mBrec$ (in particular setting a tighter cut
on the electron mode). 
Last but not least, we stress that a precise description of the $q^2$-dependence of the 
non-radiative amplitude, including short- and long- 
distance terms,
 is a key ingredient to obtain the $\ORD(\al)$ corrections at the sub-percent level (cf.~remarks at the end of \SEC \ref{sec:splittres}).

In this paper, we have specifically focused on the case of neutral hadrons, which has facilitated the implementation of 
an arbitrary $q^2$-dependence in the form factor. As discussed, we expect the conclusions for the charged modes to 
be  qualitatively similar, especially as far as the interference effects between resonant and rare modes are concerned.  Moreover, the 
same outcomes ought to  hold for the $\bar B \to \bar K^* \ell^+ \ell^-$  and the $\Lambda_b \to \Lambda  \ell^+ \ell^-$ modes.

A decisive aspect is that the remaining QED corrections, due to structure dependence, which are not incorporated in \PHOTOS,  have been shown to be free of $\ln m_\ell$ enhanced factors (cf. \SEC3.4~\cite{Isidori:2020acz}). When going to the structure dependent level, which necessitates 
the introduction of new gauge invariant interpolating operators  \cite{Nabeebaccus:2022jhu} 
and or new gauge invariant 
distribution amplitudes  \cite{Beneke:2021pkl}, new  sizeable $\ln m_{K(\pi)}$-effects can 
be expected to be present for $K(\pi)$ final states. However, they would cancel in LFU ratios~\cite{BIP16,Isidori:2020acz}, along with (other) structure dependent effects, 
for the reasons mentioned above.

Putting all these ingredients together, the present analysis 
provides an important further validation that the LFU tests so far performed by the LHCb collaboration are robust with respect to LFU violations induced by QED corrections.

\paragraph{Acknowledgement} 
RZ is supported by an STFC Consolidated Grant, ST/P0000630/1.
This project has received funding from the European Research Council (ERC) under the European Union's Horizon 2020 research and innovation programme under grant agreement 833280 (FLAY), and by the Swiss National Science Foundation (SNF) under contract 200020\_204428. This work was supported by the GLUODYNAMICS project funded by the ``P2IO LabEx (ANR-10-LABX-0038)" in the framework ``Investissements d’Avenir" (ANR-11-IDEX-0003-01) managed by the Agence Nationale de la Recherche (ANR), France.

\appendix

 \section{Kinematics}
 \label{app:kin} 
 
Following Ref.~\cite{Isidori:2020acz}, we reproduce the relevant kinematic paramteristion for the  4-body decay kinematics in terms the five independent 
variables  $(\qsq , \pBbar^{2}, \cl,\ctg , \phi_\gamma)$, where $c_i \equiv \cos \theta_i$.
We start by defining the photon and the meson momenta in the $\pBbar$-RF  (indicated by an upper index $(2)$):
\begin{alignat}{2}
\label{eq:kin2}
& k^{\FRtwo} &\;=\;&  (E_\ga^{\FRtwo}  , - \cos \Tg  |\vec{k}_\ga^{\FRtwo}| , - \sin \Tg \cos \Fg  |\vec{k}_\ga^{\FRtwo}| ,
- \sin \Tg \sin \Fg  |\vec{k}_\ga^{\FRtwo}| ) \;, \nonumber \\[0.1cm]
& \pBbar^{\FRtwo} &\;=\;& ( \pBbar,0,0,0)  \;,\quad q^{\FRtwo} =  (\pBbar- p_K)^{\FRtwo} =    (\pBbar- E_K^{\FRtwo}, | \vec{p}_K^{\,\FRtwo}|,0,0  ) = 
( E_q^{\FRtwo} , | \vec{p}_K^{\,\FRtwo}|,0,0  ) \;,
 \nonumber \\[0.1cm] 
& p_K^{\FRtwo} &\;=\;& (E_K^{\FRtwo},- | \vec{p}_K^{\,\FRtwo}| ,0,0)  \;. 
\end{alignat}
Here,
\begin{alignat}{5}
\label{eq:(2)}
& E_K^{\FRtwo} &\;=\;&  \sqrt{|\vec{p}^{\;\FRtwo}_K|^2 + m_K^2}&\;=\;&   \frac{1}{2 \pBbar} ( \pBbar^2 - q^2 + m_K^2) \;, 
\quad & &   |\vec{p}_K^{\;\FRtwo}| &\;=\;&  \frac{\la^{1/2}(\pBbar^2,q^2,m_K^2)}{2 \pBbar}  \;,
\nonumber \\[0.1cm] 
&  E_\ga^{\FRtwo} &\;=\;&  
\sqrt{  |\vec{k}_\ga^{\FRtwo}|^2 + \mga^2}  &\;=\;&  \frac{1}{2 \pBbar} \left( m_B^2 - \pBbar^2 - \mga^2 \right)   \;, \quad & & |\vec{k}_\ga^{\FRtwo}| &\;=\;&  \frac{\la^{1/2}(\pBbar^2,m_B^2,\mga^2)}{2 \pBbar}  \;,
 \nonumber   \\[0.1cm] 
& E_q^{\FRtwo} &\;=\;&   \sqrt{|\vec{p}^{\;\FRtwo}_K|^2 + q^2}  
&\;=\;&  \frac{1}{2 \pBbar} ( \pBbar^2 + q^2 - m_K^2)   \;,  
\end{alignat} 
 and 
\begin{equation}
\label{eq:Kallen}
\la(s,m_1^2,m_2^2) =  (s- (m_1 -m_2)^2)(s- (m_1 +m_2)^2)\;.
\end{equation}
The lepton momenta $\lonetwo$ depend on the angle of the leptons  w.r.t to the decay axis 
in the $q$-RF, 
\begin{alignat}{2}
& \lone^{\FRtwo} &\;=\;& 
 (\ga ( E^{\FRthree}_{\lone} + \be \cos \Tl  \modveclone{\;\FRthree})   , \ga (   \be E^{\FRthree}_{\lone}+ \cos \Tl  \modveclone{\;\FRthree}), -  \modveclone{\;\FRthree} \sin \Tl ,0 )   \;, \nonumber \\[0.1cm]
& \ltwo^{\FRtwo} &\;=\;& 
 (\ga ( E^{\FRthree}_{\ltwo} - \be \cos \Tl  \modveclone{\;\FRthree})  , \ga (   \be E^{\FRthree}_{\ltwo} - \cos \Tl  
 \modveclone{\;\FRthree}) , + \modveclone{\;\FRthree} \sin \Tl ,0 )     \;, \end{alignat}
where $E_{\lonetwo}^{\FRthree}$ and $\modveclone{\;\FRthree}$ are quantities 
defined in the $q$-RF:
\begin{alignat}{5}
 & E_{\lonetwo}^{\FRthree} &\;=\;&  \sqrt{ \modveclone{\;\FRthree}^2 + m_{\lonetwo}^2} &\;=\;& \frac{1}{2 q}(q^2 + m_{\lonetwo}^2 - m_{\ltwoone}^2) \;,  \quad & & 
   \modveclone{\;\FRthree}  &\;=\;& \frac{\la^{1/2}(q^2,m_{\lone}^2,m_{\ltwo}^2)}{2 q}  \;.
\end{alignat}
 The boost velocity $\be$  and $\ga$-factor are given by
\begin{equation}
\be =  \frac{ | \vec{p}^{\;\FRtwo}_K|}{E_q^{\FRtwo}} \;,
 \quad \ga = \frac{1}{\sqrt{1- \be^2}} = \frac{ E_q^{\FRtwo}}{q} \;.
\end{equation}

\section{More Detail on the Charm Parameterisation}
\label{app:charm}

Here we give some more detail on the charm parameterisation \eqref{eq:charmused} used in this paper. 
The perturbative evaluation of  $\Delta C_9(q^2)$ reads 
\begin{equation}
\label{eq:PT}
\Delta C_9(q^2) = (C_2 + 3 C_1)  h_c(q^2) -    \frac{ \al_s }{4 \pi} \sum_{i=1,2} C_i  F_i^{(9)}(q^2) +      \ORD(\al_s^2,C_{3,6})  \;,
\end{equation}
where $h_c$ is vacuum polarisation ($\Ima [h_c(s) ] =  \frac{\pi}{3}  R(s)$ 
with $R(s) \equiv \frac{\sigma(e^+e^- \to \text{hadrons})}{\sigma(e^+e^- \to \mu^+\mu^- )}$   an experimentally 
 well-studied ratio of cross sections \cite{PDG}), reproduced in     
 \cite{LZ2014} for example, and  the second term includes $bscc$-vertex corrections not captured by 
 the first term.
 For the latter, we have adapted the notation and results from the inclusive mode $b \to s \ell \ell$  \cite{Asatrian:2001de}.  
This treatment  falls short of effects specific to the structure of the $\bar B$-  and $\bar K$-mesons.
The Wilson coefficients $C_{3,6}$ 
correspond to the penguin induced four quark operators and can be neglected for our purposes. 
The $C_2 \approx 1$  Wilson coefficient arises at tree level and $C_1 \approx - 0.15$ is   generated  
by renormalisation group running (specifically we employ 
$(C_1,C_2,C_9)(m_b) = (-0.15,1,4.035)$ as reference values). 
The combination $C_2 + 3 C_1 \approx 0.6$ is referred to as the colour suppressed contribution whereas the radiative corrections are colour enhanced and  reduce the LO corrections 
considerably. 
In our numerical analysis we use a single subtracted point at 
$ \Delta C_9(0 \GeV^2) \approx 0.27 + 0.073 i$.

We turn to the input into the dispersion integral which is the discontinuity.  
Let us clarify the approximations used. 
In the case of infinitely narrow resonances and further assuming na\"ive factorisation (NF)
one has, $(2\pi i)^{-1} \disc [\Delta C_9]_{\textrm {NF}}(s) = 
 \frac{3 \pi }{ \al^2} (C_2 + 3 C_1)  m_\Psi \Gamma_{\Psi \to \ell^+ \ell^-} \de(s- m_\Psi^2)$. 
In this limit, one can correct for NF by multiplying the amplitude by a complex number $\rho_\Psi e^{i \de_{\Psi}}$ ($\rho_\Psi \geq 0$),
\begin{equation}
\label{eq:DC9}
\frac{1}{ 2 \pi i } \disc [\Delta C_9](s)|_{ \frac{\Gamma_\Psi}{m_{\Psi}} \to 0}  =  
 \frac{3 \pi }{ \al^2} (C_2 + 3 C_1)  \,  \rho_{\Psi} e^{i \de_{\Psi}} m_{\Psi} \Gamma_{\Psi \to \ell^+ \ell^-} \de(s- m_{\Psi}^2) + \dots \;,
\end{equation}
which parameterises its deviation. Turning to the more realistic case of finite widths, 
the dispersion integral in \eqref{eq:disp} assumes the form
\begin{eqnarray}
\label{eq:other}
& &  \frac{(q^2 \mi s_0)^n}{  2 \pi i  }  \int_{\textrm{cut}}^{\infty}  \frac{ds}{ (s \mi s_0)^n }\frac{ \disc [\Delta C_9](s)}{s 
 \mi q^2 \mi i0}  = \nonumber  \\[0.1cm]  
 & & \qquad \qquad \qquad  - \frac{3 \pi}{\al^2} (C_2 \pl 3 C_1)
  \sum_{r \in \Psi} 
\left(\frac{ q^2-s_0}{m_r^2-s_0}\right)^n \frac{\rho_r e^{i \de_{r}}  m_r \Gamma_{r \to \ell^+\ell^-} }{q^2 \mi m_r^2  \pl i m_r\Gamma_r}  
+ \dots \;.
\end{eqnarray}
As previously mentioned the dots stand for neglected multi-hadron contributions which start at $q^2 = 4 m_D^2$. 
At last,  let  us comment on the status of the resonance data, the significance of writing the dispersion relation in 
  $\Delta C_9$ rather than the  amplitude 
 and parameterising without reference to the SM.
\begin{itemize}
\item 
From the branching fractions, it has been known for a long time that 
$\rho_{J/\Psi} \approx 1.38$ and $\rho_{\Psi(2S)} \approx 1.56$ and that there are sizeable corrections to NF.
By fitting the interference of the (broad) charm resonances with the 
short distance contributions,   the corresponding correction factors were found to be even larger 
and come with opposite phase $\de_\Psi \approx \pi$ as compared to  NF \cite{LZ2014} (cf. plots and tables therein).
Later, the  LHCb collaboration  \cite{LHCb:2016due} fitted the phases of the $J/\Psi$- and 
$\Psi(2S)$-resonances
which are more challenging as they are narrow. 
Qualitatively, a four fold degeneracy   $(\de_{J/\Psi},\de_{\Psi(2S)})  \approx (\pm \frac{\pi}{2} ,\pm  \frac{\pi}{2})$   (cf.~\TAB\!3 in \cite{LHCb:2016due}) emerges which indicates
a rather small interference effect since the short distance contribution is real.
\item In principle, one could have directly written a dispersion relation for the full amplitude.
However, the amplitude and $\Delta C_9$ essentially differ by the form factor $f_+(q^2)$ which 
is an analytic function with a pole and branch points above the physical region. They 
are thus both legitimate functions for a dispersion relation. Differences comes into play when  approximations are made. In our case, they differ on how we extend from the narrow resonance limit, which is not-known from first principles,  and thus, a priori, any of the two seems as good as the other. Perhaps, the one for $\Delta C_9$ is  preferable as we know that the extension is at least correct in the case of NF, which might be seen as a reasonable qualitative
starting point (cf. end of \SEC\ref{sec:splittres} for comments on the actual numerical relevance.)
\item One may parameterise without reference to NF (cf.~\EQ (3) in  \cite{LHCb:2016due})
\begin{equation}
\label{eq:DC92}
\frac{1}{ 2 \pi i }  \disc [\Delta C_9](s) =   \eta_{r} e^{i \de_{r}}   m_{r} \Gamma_{r}   \de(s- m_{r}^2)\;,
\end{equation} 
thereby avoiding reference to the SM.
Comparison with \eqref{eq:DC9} reveals 
the relation between the parameters 
\begin{equation}
\label{eq:other}
   \eta_{r} \,  \Gamma_r =  \rho_{r} \,  \frac{3 \pi}{\al^2} (C_2 + 3 C_1) \Gamma_{r \to \ell^+\ell^-}  \;.
\end{equation}
\end{itemize}

\section{Supplementary plots}
\label{app:suppplots}

\begin{figure}[!htbp]
   \begin{center}
      \begin{overpic}[width=0.48\linewidth]{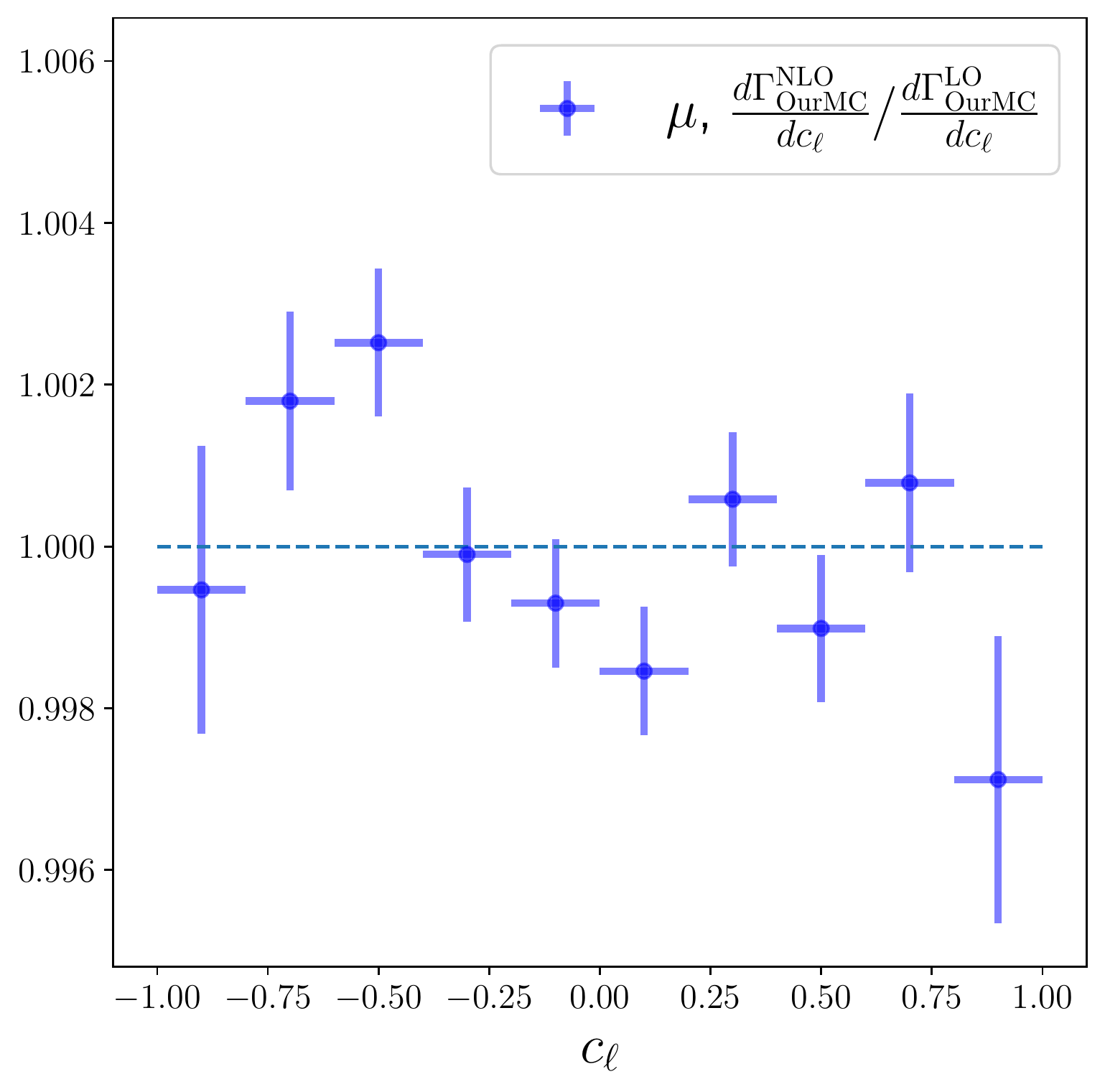}
      \put(52, 75){$\mBrec > 5.18 \GeV$}
      \end{overpic}
      \begin{overpic}[width=0.48\linewidth]{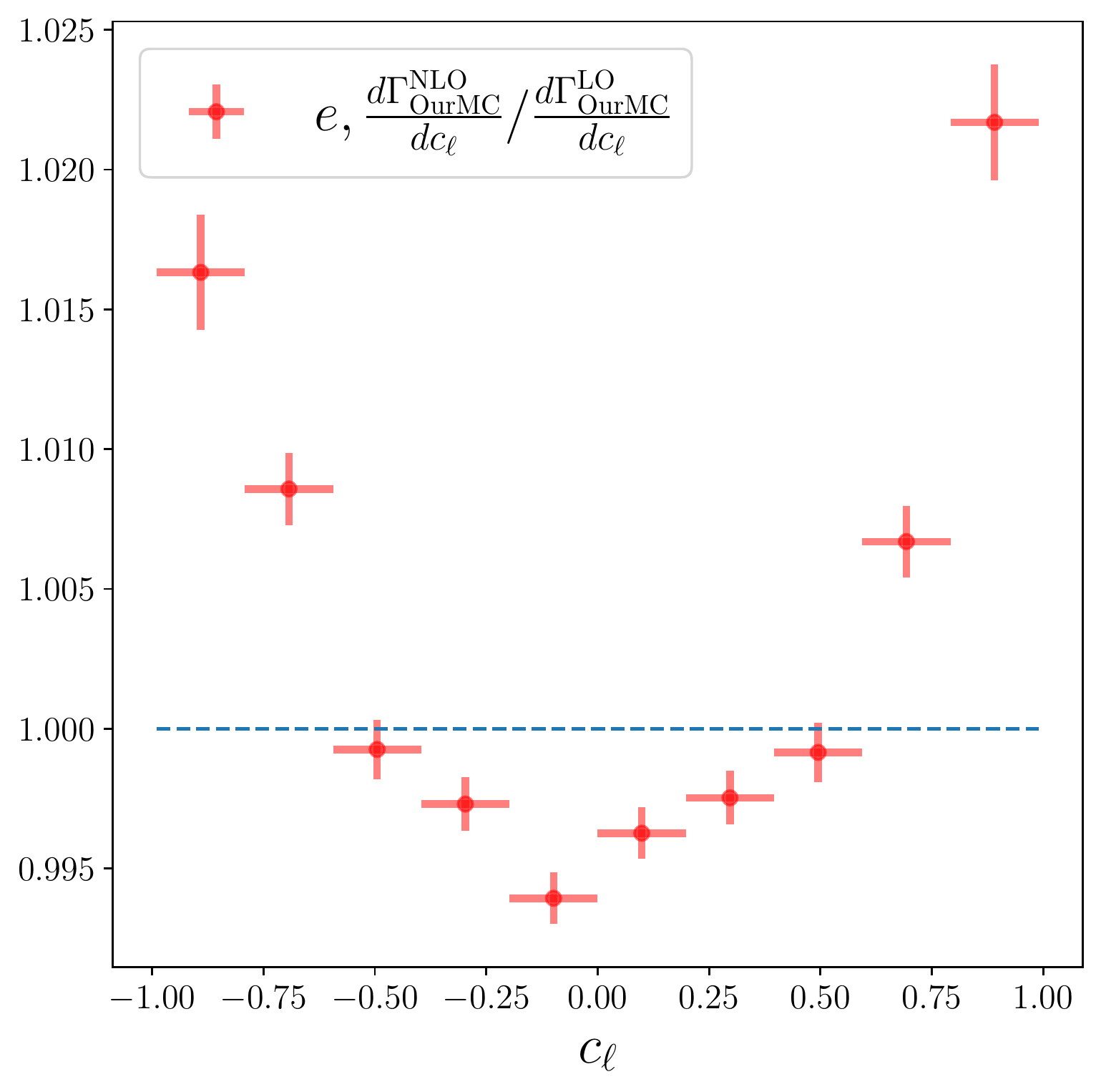}
      \put(25, 75){$\mBrec > 4.88 \GeV$}
      \end{overpic}
      \begin{overpic}[width=0.97\linewidth]{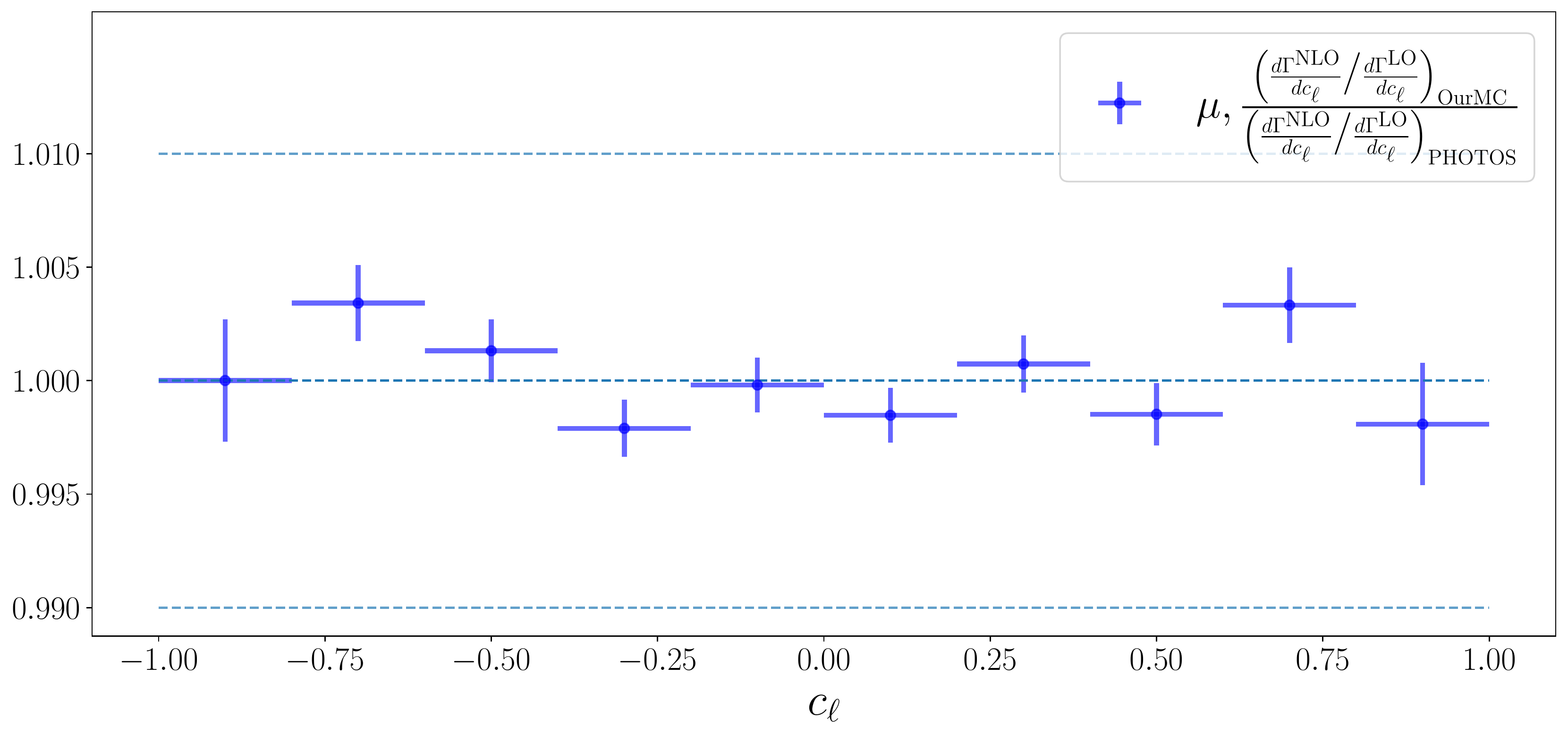}
      \put(45, 40){$\mBrec > 5.18 \GeV$}
      \end{overpic}
      \begin{overpic}[width=0.97\linewidth]{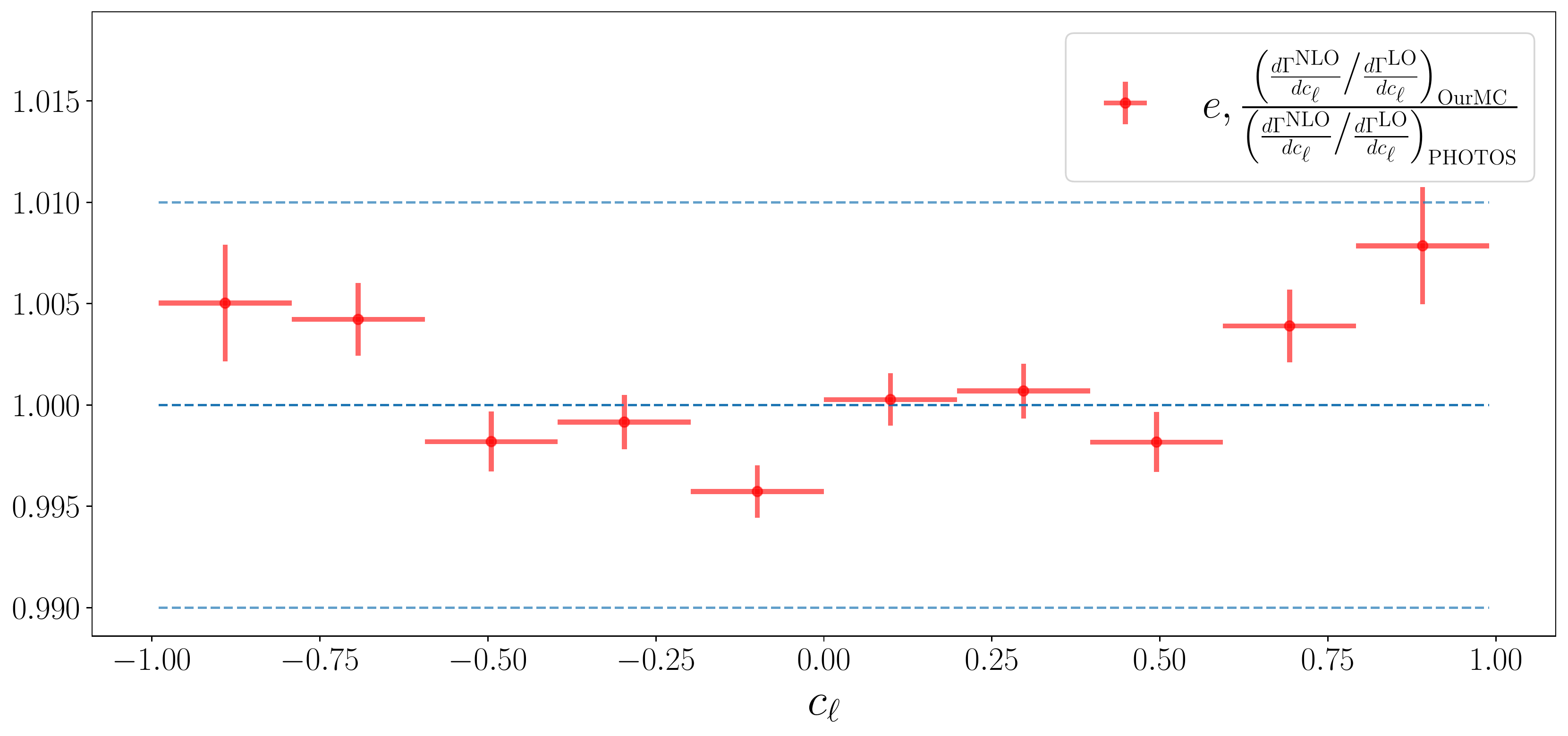}
      \put(47, 40){$\mBrec > 4.88 \GeV$}
      \end{overpic}      
	\end{center}

\caption{Short distance (form factor) plots, in the $\ctl$-variable,  
NLO over LO for  muons in blue (top left) and  for electrons in red (top right) in our MC, with appropriate cuts as in \TAB\ref{tab:mBrec}. 
The normalisation of these plots is not meaningful (cf. main text). 
However, double ratios, shown  in the middle and bottom,  of our Monte Carlo versus the \PHOTOS framework are free of ambiguities.} \label{fig:ctlcomparison}
\end{figure}

\begin{figure}[!htbp]
   \begin{center}
    \hspace{-7.8cm}   \begin{overpic}[width=0.5\linewidth]{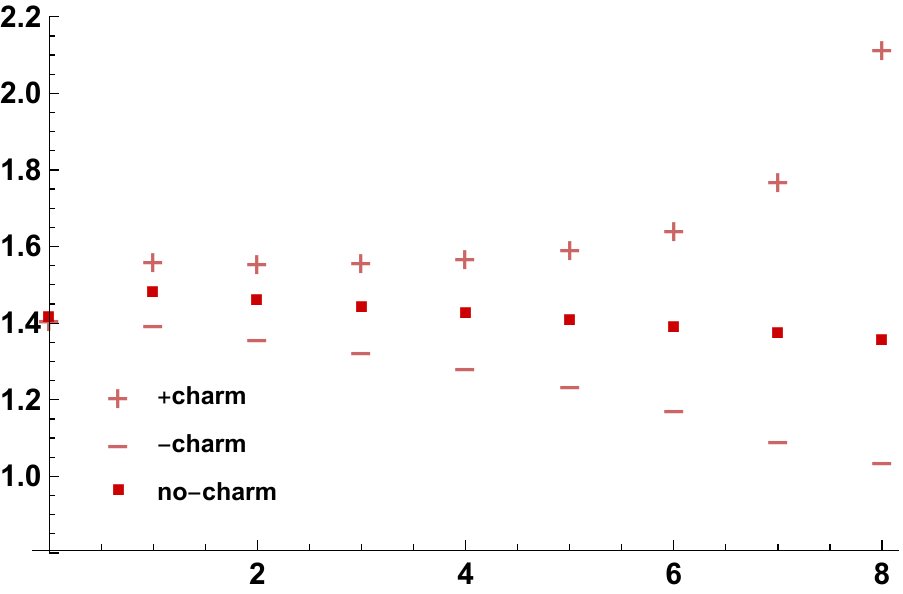} 
       \put(11,57){${ \tiny \boxed{e, \text{10}^8\text{GeV} {\frac{ d\Gamma^\textrm{NLO}}{dq^2}}} }$}
     \put(50, 60){${\scriptstyle{ \mBrec >4.88 \text{GeV} }  }$}
       \put(50, 54){${ \scriptstyle{ \delta_{J/\Psi,\Psi(2S)}=0,\pi}}$}  
       \put(50, 48){$ \scriptstyle{\text{without } |A_{\bar B \to \bar K (\Psi \to ee) }|^2}$}  
       \put(77,-5){${ \small q^2[\GeV^2] } $}
      	\hspace{7.8cm}
	\begin{overpic}[width=0.5\linewidth]{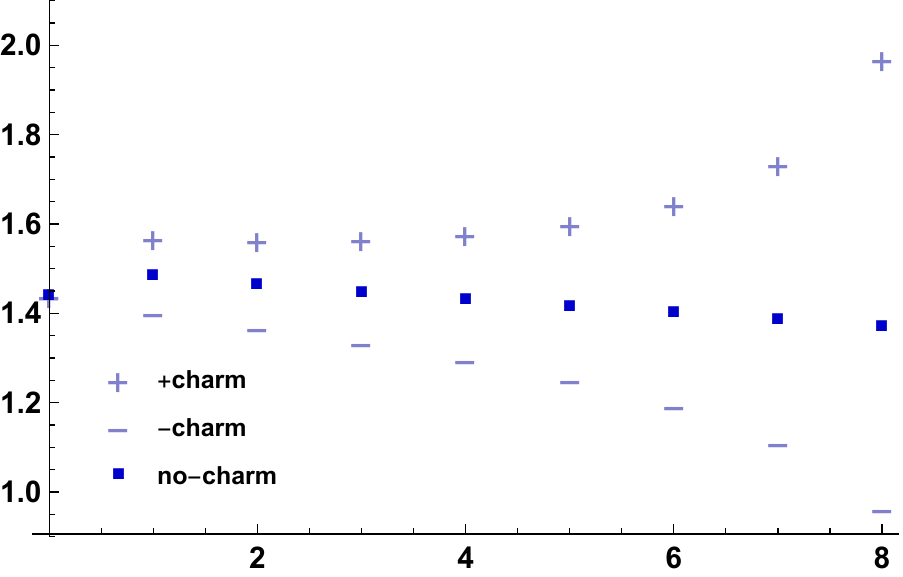} 
      \put(11,57){${ \tiny \boxed{\mu, \text{10}^8\text{GeV} {\frac{ d\Gamma^\textrm{NLO}}{dq^2}}} }$} 
       \put(50, 60){${\scriptstyle{ \mBrec >5.18 \text{GeV} }  }$}
       \put(50, 54){${ \scriptstyle{ \delta_{J/\Psi,\Psi(2S)}=0,\pi}}$}  
       \put(50, 48){$ \scriptstyle{\text{without } |A_{\bar B \to \bar K (\Psi \to \mu\mu) }|^2}$} 
       \put(77,-5){${ \small q^2[\GeV^2]  }$} 
   	\end{overpic}       
      \end{overpic}            
	\end{center}
\caption{\small  Same plots as in \FIG\ref{fig:sa_window} without LO-normalisation.}
	\label{fig:sa_window_abs}
\end{figure}

\begin{figure}[!htbp]
   \begin{center}
    \hspace{-7.8cm}   \begin{overpic}[width=0.5\linewidth]{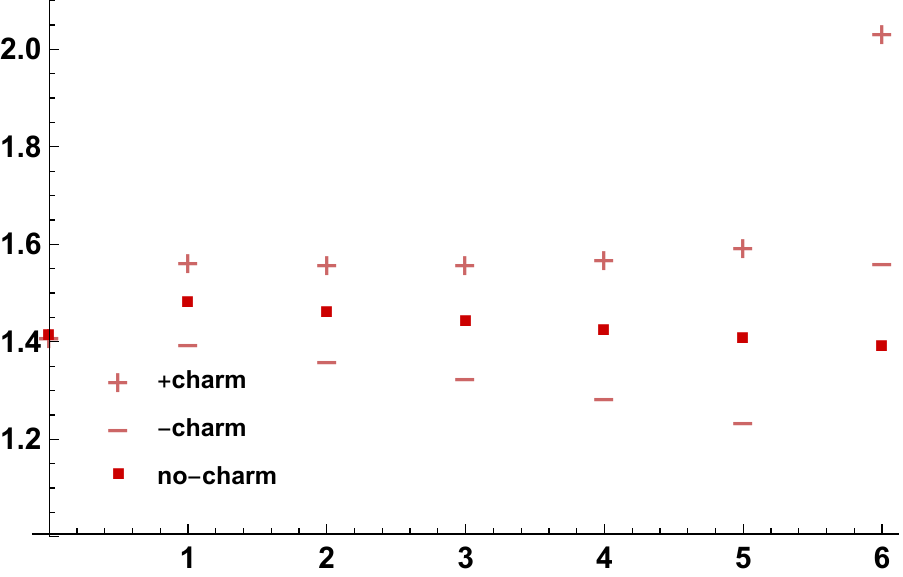} 
     { \put(11,57){${ \tiny \boxed{e, \text{10}^8\text{GeV} {\frac{ d\Gamma^\textrm{NLO}}{dq^2}}} }$}
       \put(51, 60){${\scriptstyle{ \mBrec >4.88 \text{GeV} }  }$}
    \put(51, 54){${ \scriptstyle{ \delta_{J/\Psi,\Psi(2S)}=0,\pi}}$}  
       \put(51, 48){$ \scriptstyle{\text{with }|A_{\bar B \to \bar K (\Psi \to ee) }|^2}$ }  
       \put(77,-5){${ \small q^2[\GeV^2] } $}
       }
      	\hspace{7.8cm}	\begin{overpic}[width=0.5\linewidth]{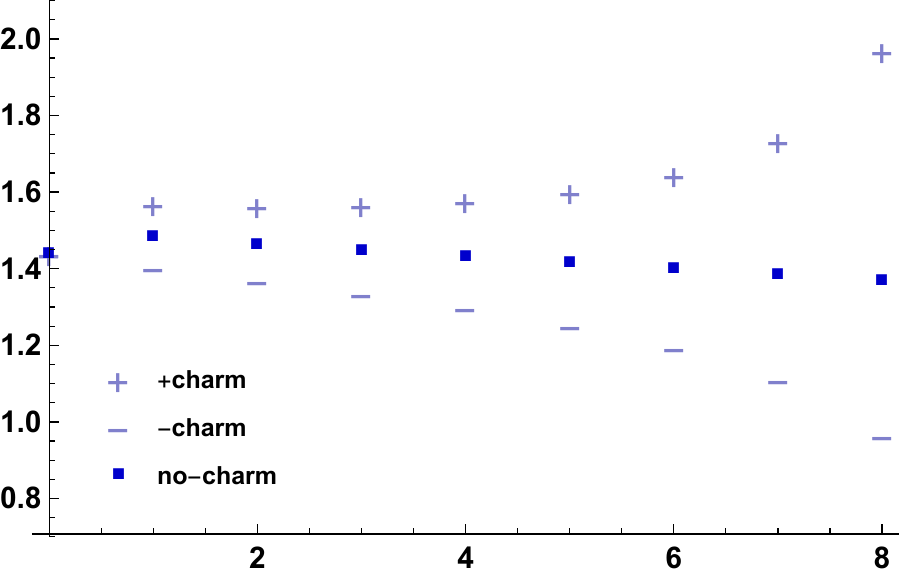} 
	\put(11,57){${ \tiny \boxed{\mu, \text{10}^8\text{GeV} {\frac{ d\Gamma^\textrm{NLO}}{dq^2}}}}$}
      \put(50, 60){${\scriptstyle{ \mBrec >5.18 \text{GeV} }  }$}
       \put(50, 54){${ \scriptstyle{ \delta_{J/\Psi,\Psi(2S)}=0,\pi}}$}  
       \put(50, 48){$ \scriptstyle{\text{with }|A_{\bar B \to \bar K (\Psi \to ee)} |^2}$}  
       \put(77,-5){${ \small q^2[\GeV^2] } $}
   	\end{overpic}       
      \end{overpic}            
	\end{center}
\caption{\small  Same plots as in \FIG\ref{fig:sa_nowindow} without LO-normalisation.}
	\label{fig:sa_nowindow_abs}

\end{figure}

In this appendix, we provide a few supplementary plots which might be of interest to certain readers of the paper.  This includes the MC-plots  in \FIGs\ref{fig:ctlcomparison} about angular distributions. The fact that QED gives rise to qualitatively different angular distribution (or higher moments) was pointed out in \cite{Gratrex:2015hna} (cf. \SEC 5) and advertised as a way to measure pure QED effects. 
Auxiliary plots for the semi-analytic approach are shown in \FIGs \ref{fig:sa_window_abs} and 
\ref{fig:sa_nowindow_abs} which can help to better understand the normalised figures since the normalisation depends on the charm input.

\section{Values of $ \fth $ used in the Monte Carlo Simulations}
\label{app:fthtable}

\TAB \ref{tab:fth} gives the values of $ \frac{\Gamma_{3}}{\Gamma_{\mathrm{tree}}} $ for various cases, which are used to calculate $ \fth $, needed for the normalisation of the Monte Carlo.

\begin{table}[h]
	\centering
	\begin{tabular}{ |c|c|c|c| }
		\hline
		Resonance & Frame for $\egammacut{i}$  & Lepton & $\frac{\Gamma_{3}}{\Gamma_{\mathrm{tree}}}$ \\ \hline\hline
		\multirow{6}{*}{Off} & \multirow{3}{*}{ $\pBbar$} & $ \mu $ & $1+\frac{\alpha}{\pi}\left( 5.1479+10.924 \ln \egammacut{\pBbar} \right)$\\
		\cline{3-4}
		& & $\ellint$ & $1+\frac{\alpha}{\pi}\left( 7.8551+23.012 \ln \egammacut{\pBbar} \right)$\\
		\cline{3-4}
		& & $e$ & $1+\frac{\alpha}{\pi}\left( 9.9605+32.222 \ln \egammacut{\pBbar} \right)$\\
		\cline{2-4}
		& \multirow{3}{*}{$ q_0 $} & $ \mu $ & $1+\frac{\alpha}{\pi}\left(6.8903+10.924 \ln \egammacut{q_0} \right)$\\
		\cline{3-4}
		& & $\ellint$ & $1+\frac{\alpha}{\pi}\left( 12.473+23.012 \ln \egammacut{q_0} \right)$\\
		\cline{3-4}
		& & $e$ & $1+\frac{\alpha}{\pi}\left( 16.772+32.222 \ln \egammacut{q_0} \right)$\\
		\hline
		\multirow{2}{*}{On, $  \delta _{J/ \psi }=0 $} &  \multirow{2}{*}{$q_0$}  & $ \mu $ & $1+\frac{\alpha}{\pi}\left(8.7111 +10.140 \ln \egammacut{q_0} \right)$\\ \cline{3-4}
		& & $\ellint$ & $1+\frac{\alpha}{\pi}\left( 16.658+22.223 \ln \egammacut{q_0} \right)$\\
		\cline{1-4}
		\multirow{2}{*}{On, $  \delta _{J/ \psi }=1.47 $} & \multirow{2}{*}{$ q_0 $} & $ \mu $ & $1+\frac{\alpha}{\pi}\left( 9.6159+9.5767 \ln \egammacut{q_0} \right)$\\
		\cline{3-4}
		& & $\ellint$ & $1+\frac{\alpha}{\pi}\left( 19.271+21.647 \ln \egammacut{q_0} \right)$\\
		\hline
	\end{tabular}
	\caption{\small $\fth$ for different cases. When the resonance is ``on", the interference of the Breit-Wigner term of the $ J/\psi $ and the rare mode is included (but not the square of the Breit-Wigner term). $ \ellint $ is a `fake' lepton that has an `intermediate' mass (between the muon and the electron), and we take $ m_{\ellint}=10\,m_{e} $. When the resonance is ``off" (ie. only rare mode), the full range of $ q_0^2 $ is integrated over. When $  \delta _{J/\psi}=0 $, the $ q_0^2 $ integration is restricted in the region $ q_0^2 \leq 9.5905 \GeV^2$, whereas when $  \delta _{J/\psi}=1.47 $, the $ q_0^2 $ integration is restricted in the region $ q_0^2 \leq 9.585 \GeV^2$. $\egammacut{i}$ is given in GeV units.
	}
	\label{tab:fth}
\end{table}

When the resonance is ``off", only the contribution from the rare mode is considered, and the integration for the total rates is performed over the full range of $ q_0^2 $. In this case, we consider two possible frames ($ \pBbar $ and $ q_0 $) for imposing the spurious cut on the photon energy $ \egamma{i} $ in order to separate 3- and 4-body events. We note that the coefficient of the soft log ($ \ln \egamma{i} $) is the same for each case, as expected. The results are given for 3 different leptons ($  \mu  $, $ \ellint $ and $ e $) for each photon energy cut.  The shorthand $ \ellint $ denotes an ``intermediate" lepton which has a mass of $ m_{\ellint}=10\,m_{e} $, which is roughly in between the muon and electron mass.

When the resonance is ``on", the contribution from the interference of the rare mode with the Breit-Wigner term of the $ J/\psi $ is included, but not the square of the Breit-Wigner term itself. In this case, we consider two possible values for the phase of the $ J/\psi $: $  \delta _{J/\psi}=0 $ (maximum interference) and $  \delta _{J/ \psi }=1.47 $ (LHCb value from \cite{LHCb:2016due}). The corresponding restrictions on the range of the $ q_0^2 $ integration are $ q_0^2\leq 9.5905 \GeV^2 $ and $ q_0^2\leq 9.585 \GeV^2 $ respectively. This is done in order to capture the maximum effect from the interference term in each case. {Only results for photon energy cuts in the $ q_0 $-RF are given, since in the $ \pBbar$-RF, the MC has extremely low efficiency.} This is because for a photon energy cut-off in the $\pBbar$-RF, the sampling for the MC also has to be performed in $\pBbar$. Then, applying a cut in $q_0^2$ (which is now a function of several sampling variables) becomes problematic, and significantly decreases the sampling efficiency. Furthermore, having an electron mass also significantly decreases the efficiency of the Monte Carlo, so we restrict ourselves to a muon mass $ m_{ \mu } $ and an ``intermediate" mass $ m_{\ellint}  (= 10 m_e)$ when interference effects are considered.

\bibliographystyle{utphys}
\bibliography{../../Refs-dropbox/References_QED.bib}

\end{document}